\documentclass[12pt]{iopart}
\usepackage{iopams}  
\usepackage{graphicx}
\usepackage{dcolumn}
\usepackage{bm}
\usepackage[utf8]{inputenc}
\usepackage[T1]{fontenc}
\usepackage{mathptmx} 
\usepackage{tikz-feynman}
\usepackage{float}
\usepackage{longtable}

\usepackage{subfigure}

\usepackage{todonotes}

\def \hhstroke{ 
    \vcenter{\hbox{
    \begin{tikzpicture}
      \begin{feynman}
        \vertex (a);
        \vertex [right=1.5cm of a] (b);
    
        \diagram* {
            (a) -- [fermion', thick] (b)
            };
        \end{feynman}
    \end{tikzpicture}}} 
}

\def \vvstrokeHKNS{ 
    \vcenter{\hbox{
    \begin{tikzpicture}
      \begin{feynman}
        \vertex (a);
        \vertex [right=1.5cm of a] (b);
    
        \diagram* {
            (a) -- [boson', thick] (b)
            };
        \end{feynman}
    \end{tikzpicture}}} 
}

\def \lineonly{ 
    \vcenter{\hbox{
    \begin{tikzpicture}
      \begin{feynman}
        \vertex (a);
        \vertex [right=1.5cm of a] (b);
    
        \diagram* {
            (a) -- [fermion, thick] (b)
            };
        \end{feynman}
    \end{tikzpicture}}} 
}

\def \lineonlyboson{ 
    \vcenter{\hbox{
    \begin{tikzpicture}
      \begin{feynman}
        \vertex (a);
        \vertex [right=1.5cm of a] (b);
    
        \diagram* {
            (a) -- [boson, thick] (b)
            };
        \end{feynman}
    \end{tikzpicture}}} 
}

\def \hhstrokeFirst{
    \vcenter{\hbox{
    \begin{tikzpicture}
      \begin{feynman}
        \vertex (a);
        \vertex [right=0.8cm of a] (b);
        \vertex [right=of b] (c);
        \vertex [right=0.8cm of c] (e);
    
        \diagram* {
            (a) -- [fermion, thick] (b),
            (c) -- [fermion', half right] (b),
            (b) -- [fermion, half right] (c), 
            (c) -- [fermionZerowidth''] (e),
            };
        \end{feynman}
    \end{tikzpicture}}} 
}
\def \hhstrokeSecond{
    \vcenter{\hbox{
    \begin{tikzpicture}
      \begin{feynman}
        \vertex (a);
        \vertex [right=0.8cm of a] (b);
        \vertex [right=of b] (c);
        \vertex [right=0.8cm of c] (e);
    
        \diagram* {
            (a) -- [fermion] (b),
            (c) -- [fermion', half right] (b),
            (b) -- [->, boson, half right, thick] (c), 
            (c) -- [fermionZerowidth''] (e),
            };
        \end{feynman}
    \end{tikzpicture}}} 
}

\def \vvstroke{
    \vcenter{\hbox{
    \begin{tikzpicture}
      \begin{feynman}
        \vertex (a);
        \vertex [right=0.8cm of a] (b);
        \vertex [right=of b] (c);
        \vertex [right=0.8cm of c] (e);

        \diagram* {
            (a) -- [Boson] (b),
            (c) -- [boson', half right] (b),
            (b) -- [Boson, half right, looseness=1.5] (c), 
            (c) -- [bosonZerowidth'] (e),
            };
        \end{feynman}
    \end{tikzpicture}}} 
}

\def \hhstrokeSeconddetail{ 
    \vcenter{\hbox{
    \begin{tikzpicture}
      \begin{feynman}
        \vertex (a);
        \vertex [right=0.8cm of a] (b);
        \vertex [right=of b] (c);
        \vertex [right=0.8cm of c] (e);
        \vertex [below=0.4cm of b, black] (b2) {\(a\)};
        \vertex [below=0.3cm of c, black] (c2) {\(b\)};
    
        \diagram* {

            (a) -- [fermion, momentum={\(p\)}] (b),
            (b) -- [antifermion', half left, reversed momentum'={[arrow shorten=0.25]\(\omega\)}, momentum={[arrow shorten=0.35]\(p-k\)}] (c),
            (b) -- [Boson, half right, looseness=1.5, momentum'={[arrow shorten=0.35]\(k\)}] (c), 
            (c) -- [fermionZerowidth'', momentum={\(p\)}] (e),
            };
        \end{feynman}
    \end{tikzpicture}}} 
}

\tikzfeynmanset{ SB/.style = { 
   decoration={markings,
        mark=at position 0.6 with {\arrow[semithick]{Stealth[black,width=3mm,length=3mm]}},
        mark=at position 0.8 with {\arrow[semithick]{Bar[black,width=3mm,length=0mm]};}},
     line width=0.3mm, 
   postaction=decorate}
}

\tikzfeynmanset{ fermion'/.style = { 
   decoration={
     markings,
     mark=at position 0.7
          with {\arrow[xshift=2mm]{Bar[black,width=3mm,length=0mm]}}
     },
     line width=0.3mm, 
   postaction=decorate}
}

\tikzfeynmanset{ fermion/.style = {
     line width=0.3mm,
   postaction=decorate}
}

\tikzfeynmanset{ boson'/.style = {boson, 
   decoration={
     markings,
     mark=at position 0.7
          with {\arrow[xshift=2mm]{Bar[black,width=3mm,length=0mm]}}
     },
     line width=0.3mm, 
   postaction=decorate}
}

\tikzfeynmanset{ antiboson'/.style = {boson, 
   decoration={
     markings,
     mark=at position 0.2
          with {\arrow[xshift=0mm]{Bar[black,width=3mm,length=0mm]}}
     },
     line width=0.3mm, 
   postaction=decorate}
}

\tikzfeynmanset{ bosonZerowidth'/.style = {boson, 
   decoration={
     markings,
     mark=at position 0.05
          with {\arrow[xshift=2mm]{Bar[black,width=3mm,length=0mm]}}
     },
     line width=0.3mm, 
   postaction=decorate}
}

\tikzfeynmanset{ antifermion'/.style = { 
   decoration={
     markings,
     mark=at position 0.2
          with {\arrow[xshift=0mm]{Bar[black,width=3mm,length=0mm]}}
     },
     line width=0.3mm, 
   postaction=decorate}
}

\tikzfeynmanset{ Boson/.style = {boson, 
     line width=0.3mm,
   postaction=decorate}
}

\tikzfeynmanset{ fermionZerowidth''/.style = { 
   decoration={
     markings,
     mark=at position 0.05
          with {\arrow[xshift=2mm]{Bar[black,width=3mm,length=0mm]}}
     },
     line width=0.3mm, 
   postaction=decorate}
}

\tikzfeynmanset{ fermionZerowidth'''/.style = { 
   decoration={
     markings,
     mark=at position 0.05
          with {\arrow[xshift=2.5mm]{Bar[black,width=3mm,length=0mm]}}
     },
     line width=0.3mm, 
   postaction=decorate}
}

\tikzfeynmanset{ boson''/.style = {boson, 
   decoration={markings,
        mark=at position 0.6 with {\arrow[semithick]{Stealth[black,width=3mm,length=3mm]}},
        mark=at position 0.8 with {\arrow[semithick]{Bar[black,width=3mm,length=0mm]};}},
     line width=0.3mm, 
   postaction=decorate}
}

\begin{document}

\title[Renormalization group analysis of a self-organized system]{Renormalization group analysis of a self-organized critical system: Intrinsic anisotropy vs random environment.}

\author{N.~V.~Antonov,$^{1, 2}$ P.~I.~Kakin,$^{1}$ 
N.~M.~Lebedev$^{2}$ and A.~Yu.~Luchin.$^{1}$ }

\address{$^1$ Department of Physics, Saint Petersburg State University,
7/9 Universitetskaya Naberezhnaya, Saint Petersburg 199034, Russia \\
$^2$ N.~N.~Bogoliubov Laboratory of Theoretical Physics, Joint Institute for Nuclear Research, Dubna 141980, Moscow Region, Russia}

\ead{n.antonov@spbu.ru, p.kakin@spbu.ru, nikita.m.lebedev@gmail.com, luhsah@mail.ru}
\vspace{10pt}

\begin{abstract}
We study a self-organized critical system coupled to an isotropic random fluid environment. The former is described by a strongly anisotropic continuous (coarse-grained) model introduced by Hwa and Kardar [Phys. Rev. Lett. {\bf 62} 1813 (1989); Phys. Rev. A {\bf 45} 7002 (1992)]; 
the latter is described by the stirred Navier--Stokes equation due to  Forster, Nelson and Stephen [Phys. Rev. A {\bf 16} 732 (1977)]. 
The full problem of two coupled stochastic equations is 
represented as a field theoretic model, which is shown to be 
multiplicatively renormalizable. The corresponding renormalization group equations possess a semi-infinite curve of fixed points 
in the four-dimensional space of the model parameters. 
The whole curve is infrared attractive for realistic values of parameters; its endpoint corresponds to the purely isotropic
regime where the original Hwa-Kardar nonlinearity becomes irrelevant. There, one is left with a simple advection of a passive scalar field by the external environment. The main critical dimensions are calculated to the leading one-loop order (first terms in the $\epsilon=4-d$ expansion); some of them are appear to be exact in all orders.
They remain the same along that curve, which makes it reasonable to interpret it as a single universality class. However, the correction exponents do vary along the curve. It is therefore not clear whether the curve survives in all orders of the renormalization group expansion or shrinks to a single point when the higher-order corrections are taken into account. 
\end{abstract}


\section{\label{sec:level1} Introduction}

It has long been observed that numerous cooperative many-body systems 
achieve a kind of critical state in the natural course of their  
intrinsic dynamics. Such systems are termed to display self-organized criticality (SOC)~\cite{Bak} and considered to stand in contrast with
more common systems, nearly-equilibrium \cite{Amit}~--~\cite{Book3} as well as strongly non-equilibrium \cite{Hinrichsen,Henkel} ones,
that need to be ``fine-tuned'' to arrive at their critical points.
 
Systems with SOC are ubiquitous in Nature and are widely encountered beyond the realm of physics in a narrow sense of the word~\cite{BTW}~--~\cite{autism}. 
Numerous examples 
are provided by biological systems~\cite{bio1,bio2}, including neural systems among them~\cite{neu1}~--~\cite{neu6}, social networks~\cite{net1}~--~\cite{net5}, and many others \cite{crop,autism}. That is why any further research of SOC remains  actual in view of a large constantly growing amount of examples and applications already known.

In original pioneering studies, SOC phenomena were described by certain speculative or conjectural models, both discrete in time and space; see, {\it e.g.} \cite{BTW}~--~\cite{BTW2}.

However, in the theory of much more common equilibrium critical behaviour, it has long being realized that discrete models (like the Ising and Heisenberg models of magnets) can be substituted with the continuous $\phi^4$ field theory, as long as their critical behaviour is concerned \cite{Amit}~--~\cite{Book3}. Thus, one can hope that the passage to effective continuous models of SOC also would not  ``throw out the child along with the bathwater.''

Such kind of model for SOC was proposed by Hwa and Kardar in \cite{HK,HK1};  
it is a continuous (coarse-grained) stochastic differential equation, designed on the basis of simplicity, dimensionality, conservation 
and symmetry considerations.

Existing experience with nearly-critical systems has shown that they are extremely sensitive to various kinds of external disturbances: inclusion of impurities, gravity, effects of environment~\cite{Onuki2}~--~\cite{AHH} and so on. In practice, such systems can hardly be isolated from the influence of surrounding medium, like turbulent motion in the atmosphere, in ocean or, especially, in forest fires.

Moreover, it was observed that impressive macroscopic manifestations of underlying microscopic dynamics of strongly interacting many-body systems (such as formation of snow-flake structures, stochastic resonance {\it etc.}) may result from such a rivalry between intrinsic dynamics and various external disturbances that, in their turn, reveal themselves as a kind of effective external friction \cite{Parisi}~--~\cite{Parisi1}.

In most common and ordinary situations, the original dynamics has some symmetries (dimensional, translation, rotational or Galilean invariance), violated by the initial or boundary conditions, specific geometry of an experimental setup and other various external perturbations.

In this paper, we consider the situation which, in a sense, is opposite: 
we study the \textit{strongly anisotropic} Hwa-Kardar (HK) model \cite{HK,HK1} of SOC coupled to the \textit{isotropic} random environment described by the stochastic Navier--Stokes (NS) equation with a random stirring force due to Forster,  Nelson, and Stephen~\cite{FNS}. 

Some explanations are here in order.
In the most intuitive and pictorial interpretation, the HK model describes a box filled with sand, with an open wall through which the sand is removed, and subjected to a ``rain of grains'' from above. The system is supposed to achieve a SOC steady state. There, the anisotropy of the external setup in the resulting dynamic model is conveyed (``delegated'') to the effective differential equation and therefore can be viewed as ``intrinsic'' or ``local'' one.

Thus, the original HK equation for the height field involves a  
distinguished direction from the very beginning and, therefore, it is strongly anisotropic due to its very formulation.
Here, the anisotropy is termed ``strong''  in the sense that one can introduce two independent canonical dimensions (length scales): 
one for the chosen  direction and the other for the orthogonal subspace.
The corresponding critical dimensions are also different; 
see, {\it e.g.} \cite{WeU,WeU2} for a more detailed discussion.

In the spirit of the theory of critical state, the model can be viewed as an example  of the universality class corresponding to a variety of physical systems. Other possible representatives
may imply the presence of local anisotropy from the very beginning, due to a certain intrinsic microscopic dynamics. This also justifies interest to the model for general spatial dimension $d$.

In contrast to those basic dynamics, the external perturbation is represented by an isotropic equilibrium random environment. For the latter, we employ the stochastic NS equation with an external stirring force introduced and studied in~\cite{FNS}.

In an earlier work  \cite{WeU,WeU2}, the velocity was  modelled by the ``rapid-change'' Kazantsev--Kraichnan velocity ensemble \cite{FGV}. It was shown  that such advection  ``washes away'' the nonlinearity of the original Hwa-Kardar equation \cite{HK,HK1} for the most realistic values of the model parameters \cite{WeU,WeU2}. 

At the same time, in some asymptotic regimes, the coupling of anisotropic dynamics with isotropic environment results in a kind of ``dimensional transmutation''\footnote{Not to be confused with the well-known mechanism of mass generation in a massless field theory.}, {\it i.e.} a certain dimensionless parameter acquires nontrivial canonical and critical dimensions; what is more, the very notion of critical dimension should be revisited; see \cite{WeU,WeU2} for detailed explanation.

Thus, it is tempting to model the advection by more realistic velocity fields governed by the stochastic dynamical NS equation rather than 
somewhat artificial Gaussian ensemble.

The full-fledged HK+NS model of the two coupled stochastic equations for the height and the velocity fields can be reformulated as a field theoretic model, which is shown to be  multiplicatively renormalizable. 

This allows one to use the well-established techniques of the quantum 
field theory to identify the possible types of critical behaviour, associated with infrared (IR) attractive fixed points of the renormalization group (RG) equations,
to calculate critical dimensions of the fields and parameters,
to find their regions of stability and to discuss probable crossover phenomena. 

It should be noted that the HK height field is ``passive'' in the sense that it has no feedback on the velocity statistics. Thus, the field of competition between various ingredients of the composite model lies in the dynamics of the height field, while the velocity dynamics remains intact.

Despite the fact that the more symmetric perturbation appears in this sense superior, the passive scalar dynamics can resist, and, in some cases, it can retain its native strong anisotropy, or, in other cases, some kind of a compromise is achieved: the resulting state is anisotropic, but no separate dimensions in the two subspaces can be introduced.

The plan of the paper is the following. In Section~\ref{sec:levelM},
we give a detailed description of the stochastic problem,
while its field theoretic formulation is presented in Section~\ref{sec:level2}.

In Section~\ref{sec:level3}, we demonstrate multiplicative renormalizability of the model by using the dimensionality and symmetry considerations. 

The practical calculations of the renormalization constants are performed in the leading one-loop approximation; they are presented in Section~\ref{sect5}. 

The RG functions and the analysis of the RG flow attractors are given in Section~\ref{sec:levelA}.

The full model involves four coupling constants. In their four-dimensional space, there is a two-dimensional surface of trivial (Gaussian) fixed points, IR attractive for $\epsilon< 0$, where $\epsilon=4-d$ is the deviation of the spatial dimension $d$ from its logarithmic value $d=4$.

There are also a few fixed points which are unstable for all $\epsilon$.
Among them there is a point in which the NS nonlinearity (and therefore the advection) becomes irrelevant, and the pure HK model is restored. In this regime, the ratios of certain kinematic coefficients acquire nontrivial dimension due to the transmutation mechanism \cite{WeU,WeU2}, while the longitudinal and transverse directions acquire independent dimensions, like in the original HK model. However, in our model this interesting asymptotic regime cannot materialize due to the instability of the corresponding fixed point.

Thus, the most interesting attractor appears to be a semi-infinite curve  
of nontrivial fixed points, where both the NS and the HK nonlinearities are simultaneously relevant. The curve is IR attractive for $\epsilon>0$; its endpoint corresponds to the regime where the HK nonlinearity becomes irrelevant and one is left with the simple passive scalar advection. Although this regime is anisotropic, no independent canonical dimensions can be introduced for the chosen direction and the orthogonal subspace; the resulting critical dimensions of the 
longitudinal and transverse subspaces are equal, in contrast with the original HK model.

In Section~\ref{sec:CD}, the corresponding critical dimensions are presented to the first order of the expansion in $\epsilon=4-d$; some results appear to be exact to all orders. It turns out that the dimensions of the basic fields and parameters
remain the same along the curve of the fixed points. Thus, it is tempting to interpret the whole curve as a single universality class. On the other hand, some correction exponents (eigenvalues of the stability matrix) do vary along the curve, which opens the possibility to interpret it as a family of classes parameterized 
by a point on the curve.
The discussion of the results and further perspectives are given in the concluding Section~\ref{sec:Conc}.

\section{\label{sec:levelM} Description of the model}

The HK stochastic equation introduced in~\cite{HK,HK1}
is a semi-phenomenological continuous model for the SOC behaviour in a coarse-grained ``running'' sandpile.
Being of a rather general importance, the model is intuitively
best illustrated  by ``sand in a box with an open top and an open side.'' 
The system is manifestly anisotropic: the new sand entering from above drives avalanches that cause some sand to exit through the side. While the sandpile surface is considered to be flat on average, it is getting rougher with time. The surface tilt is specified by a constant unit vector ${\bf n}$: ${\bf x} = {\bf x_{\bot}} + {\bf n} x_{\parallel}$, $|{\bf n}|=1$, $({\bf x_{\bot}}{\bf n})=0$. Here and below, $x=(t,{\bf x})$ represent the time-space coordinates.

More specifically, the HK equation describes the evolution of a scalar field $h=h(t,{\bf x})$ that stands for the sandpile height deviation from its average:
\begin{equation}
\partial_{t} h= \nu_{\bot 0}\, \partial_{\bot}^{2} h + \nu_{\parallel 0}\, \partial_{\parallel}^{2} h - 
\partial_{\parallel} h^{2}/2 + f.
\label{HK}
\end{equation}
Here $\nu_{\bot 0}$ and $\nu_{\parallel 0}$ are kinetic coefficients; the derivatives are 
\begin{equation}
\partial_{t} = \frac{\partial}{\partial {t}}, \quad {\bf \partial}_{\perp}^{2}=({\bf \partial}_{{\perp}}{\bf \partial}_{{\perp}})=\frac{\partial}{\partial {x_{\perp i}}}\frac{\partial}{\partial {x_{\perp i}}}; \nonumber
\end{equation}
summation over repeated tensor indices is implied here and throughout the paper; index $i$ in $x_{\perp i}$ runs from 1 to $(d-1)$ with $d$ being the dimension of surface,  and $\partial_{\parallel}=({\bf n}\partial )$. The random noise $f(x)$ has zero mean and prescribed Gaussian statistics:
\begin{equation}
\langle f(x)f(x') \rangle_f = C_{0}\,
\delta(t-t')\, \delta^{(d)}({\bf x}-{\bf x}'), \quad
\label{forceC}
\end{equation}
where $C_0>0$ is a positive amplitude and brackets $\langle\dots \rangle_f$ denote averaging over the Gaussian statistics of the random noise $f$.

In this paper, we describe the environment motion by the NS stochastic differential equation for an isotropic incompressible viscous fluid with an external random stirring force \cite{FNS}:
\begin{eqnarray}
\nabla_t v_{i} = \nu_0 \partial^2 v_{i} - \partial_i \wp + \eta_i,
\label{NS}
\end{eqnarray}
where 
\begin{eqnarray}
\nabla_{t} = \partial_{t} + ({\bf v} {\bf \partial}) 
\label{nabla}
\end{eqnarray}
is the Lagrangian (Galilean covariant) derivative, $\wp$ is the pressure
and $\nu_0$ is the kinematic viscosity coefficient.
Due to incompressibility of the fluid, the velocity field is transverse:
 $({\bf  \partial} {\bf v}) = 0$, in the sense that 
 $({\bf  k} {\bf v}) = 0$ in the momentum representation.
 Thus, the pressure can be expressed in terms of ${\bf v}$ as  
 $\wp = - \partial^{-2}\partial_i\partial_k v_i v_k$. In a more explicit notation,
\begin{eqnarray} 
\wp(t, {\bf x}) = - \int d{\bf x'}\, \Delta({\bf x}-{\bf x'}) 
\partial_i\partial_k\, v_i(t, {\bf x'}) v_k (t, {\bf x'}),
\label{pressure}
\end{eqnarray}
where $\Delta({\bf x}-{\bf x'})$ is the Green function for the Laplace equation,
$\partial^2\Delta({\bf x}-{\bf x'})=\delta({\bf x}-{\bf x'})$.
 
Like the velocity field, the external random force per unit mass $\eta_i$ 
should also be taken to be transverse\footnote{There is no loss of generality: the longitudinal (potential) part of the force would be canceled by the pressure term.}. In the spirit of the pioneering paper \cite{FNS}, we choose for $\eta_i$ a Gaussian probability distribution with zero average and a given correlation function
\begin{eqnarray}
\langle \eta_i (t, {\bf x}) \eta_j (t',{\bf x}') \rangle_{\eta} = 
D_0\,\delta(t-t')\, D_{ij} ({\bf x}-{\bf x}'),
\label{forceD}
\end{eqnarray}
where the brackets $\langle \dots \rangle_{\eta}$ stand for the averaging over the noise statistics, $D_{0}>0$ is a positive amplitude factor and
\begin{eqnarray} 
D_{ij} ({\bf x}-{\bf x}')= \, \int \frac{d\mathbf{k}}{(2\pi)^d}
P_{ij} (\mathbf{k})\,  \: \mathrm{exp} \, i(\mathbf{k}(\mathbf{x-x'})) \,
\label{forceD1}
\end{eqnarray}
with the transverse projector $P_{ij}({\bf k}) = \delta_{ij} - k_i k_j / k^2$
 and the wave number  $k\equiv |{\bf k}|$.
 
The temporal $\delta$ correlation in (\ref{forceD}) is stipulated by the Galilean symmetry, while the choice (\ref{forceD1}) is probably the simplest but a very representative one in the following sense. Without the projector, it would correspond to a uniform (that is, independent of the momentum and the frequency) random stirring of the fluid at all scales ${\bf k}$. 
The presence of the projector, caused by the incompressibility of the fluid, leads to entanglement between different ${\bf k}$-modes and, simultaneously, introduces non-locality. Indeed, in the coordinate representation one obtains:
\begin{eqnarray}
D_{ij} ({\bf x}-{\bf x}') = P_{ij} ({\bf\partial})\, \delta({\bf x}-{\bf x'}) = \delta_{ij} \delta({\bf x}-{\bf x'})
- \partial_i \partial_j \, {\Delta} ({\bf x}-{\bf x'}) ,
\label{forceDD}
\end{eqnarray}
with $\Delta({\bf x}-{\bf x'})$ from (\ref{pressure}). 

It is easily seen that the correlation function (\ref{forceD}), (\ref{forceD1}),
(\ref{forceDD}) satisfies the transversality condition $\partial_i\, D_{ij}=0$.
Furthermore,  (\ref{forceD1}) involves a finite mode at ${\bf k}=0$; it can be interpreted as an overall macroscopic random ``shaking'' of the fluid container as a whole; see footnote${}^{15}$ in \cite{FNS}.

The coupling of the fields $h$ and ${\bf v}$ is introduced by the ``minimal'' substitution \mbox{$\partial_{t} \to \nabla_{t} = \partial_{t} + ({\bf v} {\bf \partial})$} in Eq.~(\ref{HK}) with the Galilean covariant derivative  $\nabla_{t}$ from Eq.~(\ref{nabla}). 

Equations (\ref{HK}), (\ref{NS}) are studied on the entire $t$ axis; the retardation condition is assumed; the asymptotic conditions for the fields $h$ and ${\bf v}$ at $t\rightarrow -\infty$ are irrelevant due to the presence of the noises. This completes formulation of the problem.

As we will show below, in the vicinity of $d=4$, the nonlinear terms in Eqs.~(\ref{HK}) and (\ref{NS}) simultaneously become logarithmic, that is, they are of the same relevance in the sense of Wilson. This {\it a posteriori} justifies the choice of the statistics of the random force $\eta_i$ that provides a full-blooded diffusion-advection problem.

\section{\label{sec:level2}Field theoretic formulation of the model} 

The full stochastic problem~(\ref{HK})--(\ref{forceD}) is equivalent (see, {\it e.g.} Sec.~5.3 in~\cite{Book3} and references therein) to the field theoretic model with the doubled set of fields $\Phi=\{h,h',{\bf v},{\bf v'}\}$ and the action functional 
\begin{eqnarray}
    S(\Phi) &= C_0 h'h'/2 + h'\{-\nabla_t h + \nu_{\parallel 0} \partial_{\parallel}^{2} h +
\nu_{\bot 0} \partial_{\bot}^{2} h - 
\partial_{\parallel} h^{2}/2\} +
\nonumber \\
&+ D_0 {\bf v'}^2/2+
{\bf v'}\{-\nabla_t{\bf v}+ \nu_0 \partial^{2} {\bf v}\}.
\label{action_ve}
\end{eqnarray}
For brevity, here and below in similar expressions, we tacitly imply the needed integration over the arguments $x=\{t,{\bf x}\}$; in particular, the first term in (\ref{action_ve}) stands for $C_0\int dt \int d{\bf x}\,h'(t,{\bf x})h'(t,{\bf x})/2$.
The pressure term is omitted owing to the transversality of the auxiliary field ${\bf v}'$ that acts as a kind of transverse projector.

In the frequency-momentum $(\omega-{\bf k})$ representation, the bare propagators for the model~(\ref{action_ve}) are:
\begin{eqnarray}
    \langle hh' \rangle_0 &= \langle h'h \rangle_0^{*} = \frac{1}{-i\omega + \epsilon({\bf k})}, \quad \
    \langle h'h' \rangle_0= 0, \quad \
    \langle hh \rangle_0 = \frac{C_0}{\omega^2+\epsilon^2({\bf k})}, \nonumber \\
    \langle v_i v'_j \rangle_0 &= \langle  v'_i v_j \rangle_0^{*} = \frac{P_{ij}({\bf k})}{-i\omega + \nu_0 k^2}, \quad
    \langle v'_i v'_j \rangle_0 = 0, \,\,\,\,\,\,
    \langle  v_i v_j \rangle_0 = \frac{D\, P_{ij}({\bf k})}{\omega^2 + {\nu_0}^2 k^4},
    \label{propagators}
\end{eqnarray}
where $\epsilon({\bf k}) = \nu_{\parallel0}{ k}_{\parallel}^2 + \nu_{\perp0}{\bf k}_{\perp}^2$ and $P_{ij}({\bf k}) = \delta_{ij} - k_i k_j / k^2$. 
We will use the following lines for the propagators when drawing Feynman's diagrams:
\begin{eqnarray}
     \langle hh' \rangle_{0} = \hhstroke, & \quad \langle hh \rangle_{0} = \lineonly, \quad
     \nonumber
     \\
     \langle v_i v'_j \rangle_{0} = \vvstrokeHKNS, & \quad \langle v_i v_j \rangle_{0} = \lineonlyboson.
     \nonumber
\end{eqnarray}

The three vertices $-h'\partial_\parallel h^2/2$, $-h'({\bf v} {\bf  \partial})h$ and $-{\bf v'}({\bf v}{\bf  \partial}){\bf v}$ correspond to the vertex factors~$ik_\parallel^{h'}$, $i k_j^{h'}$ and $i(k_n^{{\bf v'}} \delta_{sj} + k_s^{{\bf v'}} \delta_{nj})$ respectively. 
The upper indices denote the field for which this momentum is an argument (i.e., the momentum ``flows'' through the corresponding leg of the vertex). The form of the vertex factors can be explained by the fact that 
one can move the derivatives in all those vertices onto the primed fields using the integration by parts and the transversality of the field ${\bf v}$.   

There are two coupling constants $g_0$ and $w_0$ (expansion parameters in the ordinary perturbation theory):
$ g_0=C_0/\nu_{\bot0}^{3/2}\nu_{\parallel0}^{3/2} \sim {\Lambda}^{\epsilon},
\quad w_0=D_0/\nu_{0}^3\sim {\Lambda}^{\epsilon}$. 
These expressions follow from the canonical dimensions analysis (see below) and define the typical ultraviolet (UV) momentum scale $\Lambda$.

\section{\label{sec:level3}UV divergences and renormalization of the model}

We study UV divergences in the Green's functions of the model (\ref{action_ve}) using the canonical dimensions analysis (see, {\it e.g.}~\cite{Book3}, Sections~1.15 and 1.16). There are two independent scales (the time scale $T$ and the length scale $L$), so that the canonical dimension of any quantity $F$ is described by the
frequency dimension $d_{F}^{\omega}$ and the momentum dimension $d_{F}^{k}$:
$[F] \sim [T]^{-d_{F}^{\omega}} [L]^{-d_{F}^{k}}$ (see~\cite{Book3}, Sections~1.17 and 5.14).
Normalization conditions $d_{k_i}^k=-d_{x_i}^k=1,\ d_{k_i}^{\omega} =d_{x_i}^{\omega }=0,\
d_{\omega }^k=d_t^k=0,\  d_{\omega }^{\omega }=-d_t^{\omega }=1$ allow to calculate the canonical dimensions of all the fields and the parameters entering the dimensionless action functional (\ref{action_ve}). The total canonical dimension is defined as $d_{F}=d_{F}^{k}+2d_{F}^{\omega}$; it 
should be used in RG analysis of dynamic models instead of the momentum dimension (used in analysis of static models, see \cite{Book3}, Section 5.14). Canonical dimensions for the model~(\ref{action_ve}) are presented in Table~\ref{canonical dimensions}.

\begin{table}[h]
\caption{Canonical dimensions 
in the model~(\ref{action_ve}); $\epsilon= 4-d$.}
\label{canonical dimensions}
\begin{tabular}{|c||c|c|c|c|c|c|c|c|c|}
 \hline
$F$&$h(x)$&$h'(x)$&${\bf v}(x)$&${\bf v'}(x)$&$C_0$,\ $D_{0}$&$\nu_0$,\ $\nu_{\parallel0}$,\ $\nu_{\bot0}$&$g_{0}$,\ $w_0$&\mbox{$x_{10}$,\ $x_{20}$}&{$\mu$,$m$}\\
 \hline
$d^{\omega}_F$&$1$&$-1$&$1$&$-1$&$3$&$1$&$0$&$0$&$0$\\
 \hline
$d^{k}_F$&$-1$&$d+1$&$-1$&$d+1$&$-d-2$&$-2$&$\epsilon$&$0$&$1$\\
 \hline
$d_F$&$1$&$d-1$&$1$&$d-1$&$\epsilon$&$0$&$\epsilon$&$0$&$1$\\
 \hline
\end{tabular}
\end{table}
Here $m$ is the IR cut-off and $\mu$ is the renormalization mass (also referred to as normalization point or reference momentum scale), the additional parameters of the renormalized model; see below and~\cite{Book3}. 

As can be seen from Table~\ref{canonical dimensions}, the three coefficients $\nu_0$, $\nu_{\parallel0}$ and $\nu_{\bot0}$ have identical dimensions. This means that their ratios are completely dimensionless and may enter the expressions for the $\beta$  functions (see below). Thus, these ratios should be treated as additional coupling constants, although they are not expansion parameters.
One can choose $x_{1,0} = \nu_{\parallel 0}/\nu_0$ and $x_{2,0} = \nu_{\perp 0}/\nu_0$.

All these coupling constants simultaneously become dimensionless (and the respective interactions become marginal in the sense of Wilson) at $d=4$. Thus, the model as a whole is logarithmic at $d=4$, and
the role of the expansion parameter in the RG perturbation theory
is played by the single variable $\epsilon=4-d$ that  
measures the deviation from logarithmicity.

To establish renormalizability of the model~(\ref{action_ve}), let us calculate the UV divergence indices of 1-irreducible Green's functions for the model. The UV divergence index $\delta_{\Gamma}$ for a Green's function $\Gamma$ that involves $N_h$ fields $h$, $N_{h'}$ fields $h'$, etc., coincides with the total canonical dimension $d_{\Gamma}$ of that function in the frequency-momentum representation, taken at the logarithmic dimension (see, {\it e.g.} Sec.~3.2 and Sec.~5.15 in~\cite{Book3} for detailed explanation):
\begin{eqnarray}
\delta_{\Gamma} &=& d_{\Gamma}|_{d=4}=(d+2-d_h N_h-d_{h'} N_{h'}-d_v N_{{\bf v}}-d_{{\bf v'}} N_{{\bf v'}})|_{d=4}=\nonumber\\
&=& 6- N_h-3 N_{h'}- N_{{\bf v}}-3 N_{{\bf v'}}.
\end{eqnarray}

Superficial UV divergences (those that must be eliminated by the renormalization procedure) can only be present in the Green's functions with non-negative divergence index. However, the formal index $\delta_{\Gamma}$ should be adjusted: every field $h'$ and ${\bf v'}$ entering an 1-irreducible Green's function cause additional external momentum to appear as an overall factor. It happens due to the form of the vertices that allows spatial derivative to be moved onto those fields using integration by parts. Thus, the real divergence index $\delta_{\Gamma}'$ is given by the expression:
\begin{equation}
\delta_{\Gamma}'= \delta_{\Gamma} - N_{h'} - N_{{\bf v'}} =
6- N_h-4 N_{h'}- N_{{\bf v}}-4 N_{{\bf v'}}.
\end{equation}

As a manifestation of causality, all the 1-irreducible Green's functions with $N_{h'}=N_{{\bf v'}}=0$ vanish and, therefore, do not require counterterms; see, {\it e.g.}~\cite{Book3}, Section~5.4.

Thus, we arrive at the following list of possible counterterms for the model~(\ref{action_ve}): 
\setlength{\extrarowheight}{6.5pt}
\begin{longtable}{c}

$\langle  {\bf v}' {\bf v} \rangle_{\rm{1-ir}}$ \ \ $(\delta_{\Gamma}=2,\delta'_{\Gamma}=1)$ \ \ with counterterm \ \ $ {\bf v}'\partial^2 {\bf v}$,\\ 
  $\langle  {\bf v}' {\bf v} {\bf v}  \rangle_{\rm{1-ir}}$ \ \ $(\delta_{\Gamma}=1,\delta'_{\Gamma}=0)$ \ \ with counterterm\ \ $ {\bf v}'( {\bf v}\partial) {\bf v}$,\\
    $\langle  {\bf v}' {\bf v}h \rangle_{\rm{1-ir}}$ \ \ $(\delta_{\Gamma}=1,\delta'_{\Gamma}=0)$ \ \ with counterterm \ \ $( {\bf v}' {\bf v})\partial_{\parallel}h$,\\
        $\langle  {\bf v'}hh \rangle_{\rm{1-ir}}$ \ \ $(\delta_{\Gamma}=1,\delta'_{\Gamma}=0)$ \ \ with counterterm  \ \ $h( {\bf v'}\partial)h$, \\ 
  $\langle  {\bf v'}h\rangle_{\rm{1-ir}}$ \ \ $(\delta_{\Gamma}=2,\delta'_{\Gamma}=1)$ \ \ with counterterm  \ \ $( {\bf v'}\partial)h$, \\
   $\langle h'h\rangle_{\rm{1-ir}}$ \ \ $(\delta_{\Gamma}=2,\delta'_{\Gamma}=1)$ \ \ with counterterms \ \ $h'\partial_{\parallel}^2h$, $h'\partial_{\bot}^2h$,\\
    $\langle h'hh\rangle_{\rm{1-ir}}$  \ \ $(\delta_{\Gamma}=1,\delta'_{\Gamma}=0)$ \ \  with counterterm \ \ $h'\partial_{\parallel} h^2$,\\
  $\langle h'h {\bf v}\rangle_{\rm{1-ir}}$ \ \ 
    $(\delta_{\Gamma}=1,\delta'_{\Gamma}=0)$ \ \ with counterterm  \ \ $h'( {\bf v}\partial)h$, \\ 
    $\langle h' {\bf v}\rangle_{\rm{1-ir}}$ \ \ $(\delta_{\Gamma}=2,\delta'_{\Gamma}=1)$ \ \  with counterterm \ \ $( {\bf v}\partial)h'$,\\
    $\langle h' {\bf v} {\bf v} \rangle_{\rm{1-ir}}$ \ \ $(\delta_{\Gamma}=1,\delta'_{\Gamma}=0)$  \ \ with counterterm \ \ $ {\bf v}^2\partial_{\parallel} h'$.
     \label{contr_dim}
\end{longtable}
Here, the integration over $x=\{t,{\bf x}\}$ is implied. Thus,
if the difference between two counterterms is a total derivative, 
they should be treated as the same item. 
Furthermore, the counterterms $({\bf v}\partial)h'$, $( {\bf v'}\partial)h$
and $h( {\bf v'}\partial)h$
reduce to total derivatives due to the transversality of the fields ${\bf v}$ and ${\bf v'}$, vanish after the integration over ${\bf x}$ and can be ignored.

The counterterm $\langle h' \rangle_{\rm{1-ir}}$ renormalizes the average of the random noise mean $\langle  f\rangle$ directly connected to the average $\langle  h\rangle$ by the averaging of Eq.~(\ref{HK}).
It is the deviations from the averages that are of interest here,
so that all those averages can be simultaneously treated as constants and 
can be taken to be equal to zero which allows one to ignore their renormalization. This treatment is similar to the way in which the shift of the critical temperature is ignored within the RG analysis of phase transitions in field-theoretic models; see, {\it e.g.} sections~1.20 and 1.21~in~\cite{Book3}.

The remaining list of possible counterterms can be further pared down. 

The passivity of the field $h$ (which broadly means that the dynamic of the field ${\bf v}$ is not affected by the field $h$) in field-theoretic terms translates to the vanishing of the full Green's functions with $N_{h'}>0$, $N_h=0$, $N_{{\bf v}}+N_{{\bf v'}}>0$ and the 1-irreducible Green's functions with 
$N_{h}>0$, $N_{h'}=0$, $N_{{\bf v}}+N_{{\bf v'}}>0$:
no needed diagrams can be constructed.
Thus, the counterterms $({\bf v}' {\bf v})\partial_{\parallel}h$ and $h( {\bf v'}\partial)h$ should be omitted.

Furthermore, the action functional (\ref{action_ve}) is invariant with respect to the Galilean transformation
\begin{eqnarray}
&{\bf v}(t, {\bf x}) \to {\bf v}(t, {\bf x} +{\bf u}\,t) - {\bf u}, \qquad 
&{\bf v'}(t, {\bf x}) \to {\bf v'}(t, {\bf x} +{\bf u}\,t),
\nonumber \\
    &h(t,{\bf x}) \to h(t,{\bf x}+{\bf u}\,t), \quad
&h'(t,{\bf x}) \to h'(t,{\bf x}+{\bf u}\,t)  
\label{galilean}
\end{eqnarray}
with a constant vector ${\bf u}$. 
Since the derivative (\ref{nabla}) is covariant with respect to this symmetry, it can appear in the counterterms only as a single unit. 
For example, the counterterm $h'\partial_t h$ is forbidden by the real divergence index (due to  the form of the vertices, where $h'$ always
stands under a spatial derivative). Although its Galilean partner $h'({\bf v}\partial)h$ is allowed by the real index of divergence, it cannot appear by itself. Thus, the both structures are forbidden. This symmetry also excludes the counterterms $ {\bf v}'( {\bf v}\partial) {\bf v}$ and $ {\bf v}^2\partial_{\parallel} h'$. 

The Hwa-Kardar equation (\ref{HK}) has its own Galilean-type symmetry $h(t,{\bf x}) \to h(t,{\bf x}+u{\bf n}\,t)-u{\bf n}$, where $u=const$, that is broken in the full-scale model (\ref{action_ve}). 
However, it leaves its trace in some relations for the Feynman diagrams without the fields ${\bf v}$ and ${\bf v'}$. For example, that symmetry, 
were it not violated, 
would forbid the counterterm $h'\partial_{\parallel} h^2$. Thus, in the diagrams for the Green's function  $\langle h'hh\rangle_{\rm{1-ir}}$ without the fields ${\bf v}$ and ${\bf v'}$ the UV divergences must cancel each other, as can be confirmed by the practical calculation.

As a result, we arrive at the conclusion that there are only four counterterms, namely, $h'\partial_{\parallel}^2 h$, $h'\partial_{\bot}^2 h$, $h'\partial_{\parallel}h^{2}$ and $({\bf v'}\partial^2 {\bf v})$ related to the 1-irreducible functions $\langle h'h \rangle_{{\rm{1-ir}}}$, 
$\langle h'hh \rangle_{{\rm{1-ir}}}$, and 
$\langle {\bf v'}{\bf v} \rangle_{{\rm{1-ir}}}$, respectively. This means that the model (\ref{action_ve}) is renormalizable and its renormalized action reads
\begin{eqnarray}
S_R(\Phi)& = g\mu^{\epsilon}\nu_{\bot}^{3/2}\nu_{\parallel}^{3/2} h'h'/2  + h'\{-\nabla_th+ Z_1\nu_{\parallel} \partial_{\parallel}^{2} h +
Z_2\nu_{\bot} \partial_{\bot}^{2} h - Z_{4} 
\partial_{\parallel} h^{2}/2\}+ \nonumber \\
& + w\mu^{\epsilon}\nu^3{\bf v'}^2/2+ {\bf v'}\{-\nabla_t {\bf v}+ Z_3\nu \partial^{2} {\bf v}\}.
\label{act_r}
\end{eqnarray}
Here the fields were substituted with their renormalized counterparts without a change of notation (the fields ${\bf v}$ and ${\bf v'}$ are not renormalized altogether while scalar fields were substituted as $h \to Z_{h}h,$ $h' \to Z_{h'}h'$). 

The coupling constants and other parameters are related to their counterparts as follows:
\begin{eqnarray}
g_0=Z_g g\mu^{\epsilon}, \quad w_0=Z_w w\mu^{\epsilon}, \quad x_{1,0}=Z_{x_1} x_1,\quad x_{2,0}=Z_{x_2} x_2, \nonumber \\
    \nu_{0} = \nu_{} Z_{\nu} , \quad\nu_{\parallel0} = \nu_{\parallel} Z_{\nu_{\parallel}} , \quad \nu_{\perp0} = \nu_{\perp} Z_{\nu_{\perp}}.
    \label{renorm_param}
\end{eqnarray}
Here $Z_{i}$ 
are the renormalization constants; they can be expressed in terms of the constants $Z_1$, $Z_2$, $Z_{3}$, and $Z_4$:
\begin{eqnarray}
    Z_{h} = Z^{-1}_{h'}=Z_{4}, 
    \nonumber \\ Z_g = Z_{1}^{-3/2}Z_{2}^{-3/2}Z_4^2, \quad Z_w = Z_{3}^{-3}, \quad Z_{x_1} = Z_{1}Z_{3}^{-1}, \quad Z_{x_2} = Z_{2}Z_{3}^{-1}, \nonumber \\
    Z_{\nu_{\parallel}}=Z_1, \quad Z_{\nu_{\perp}}=Z_2, \quad  Z_{\nu}=Z_3
\label{ZZZ}
\end{eqnarray}
(we note that $Z_{{\bf v}}=Z_{{\bf v'}}=1$).

\section{One-loop calculations \label{sect5}} 

In this section, we present the calculation of the renormalization constants 
$Z_1$--$Z_4$ in the one-loop approximation, that is, in the leading order in the couplings $g$ and $w$. 

Let us briefly discuss the choice of the IR regularization. In the model (\ref{forceD}), the IR regularization in the diagrams is provided by the external frequencies and momenta.
However, in the simple way of calculation we adopt here, the diagrams are calculated at zero external frequencies, while the integrands are expanded in the external momenta to the desired order. Then, to ensure the IR convergence, it is necessary to introduce the cut-off of integration momenta at a certain value $k=m>0$ in expressions like 
(\ref{forceD1}). For the correlation function (\ref{forceC}) this implies the 
replacement 
\begin{equation}
\delta^{(d)}({\bf x}-{\bf x}')  \to  \int_{k>m}^{} \frac{d\mathbf{k}}{(2\pi)^d}
  \: \mathrm{exp} \, i(\mathbf{k}(\mathbf{x-x'})). \quad
  \nonumber 
\end{equation}
According to the general statement, the renormalization constants in the MS scheme do not depend on the specific choice of the IR regularization; see, {\it e.g.} Section~3.19  in~\cite{Book3}. Thus, the results we obtain here for the renormalization constants, RG functions, fixed points and critical dimensions are valid for all  $m\ge0$.

The one-loop approximation for the 1-irreducible Green's functions $\langle h h' \rangle_{\rm 1-ir}$ and $\langle  { {\bf v}} { {\bf v}}' \rangle_{\rm 1-ir}$ is:
\begin{eqnarray}
     \langle h h' \rangle_{\rm 1-ir} &=& {\rm i}\omega - \nu_{\parallel} p_{\parallel}^2 Z_1 - \nu_{\bot} p_{\bot}^2 Z_2 + \nonumber\\ &&\hhstrokeFirst \ + \hhstrokeSecond \ \ , 
    \label{hh'}
\end{eqnarray}
\vspace{-3.5ex}
\begin{equation}
     \langle  { {\bf v}} { {\bf v}}' \rangle_{\rm 1-ir} = {\rm i}\omega - \nu p^2 Z_3 \ + \vvstroke \ \ . 
     \label{vv'}
\end{equation}
Here we used diagrammatic technique introduced in Section \ref{sec:level2}. 

The first diagram in Eq. (\ref{hh'}) and the diagram in Eq. (\ref{vv'}) are calculated in the standard manner; see Appendix A in \cite{Vitalik} (calculation of diagram $D_2$) and Appendix in \cite{AKL} (calculation of diagram $A_9$). The results are
\begin{eqnarray}
  \hhstrokeFirst =   \frac{S_d }{(2\pi)^d} \nu_{\parallel} \nu_{\perp}^{\frac{4-d}{2}} p_\parallel^2 \frac{1-d}{4d}\frac{ g}{\epsilon}\,\Bigg|_{d=4}= 
  -\frac{1 }{8\pi^2} \nu_{\parallel}  p_\parallel^2 \frac{3}{16}\frac{ g}{\epsilon};\label{B3}\\
     \vvstroke = - \frac{S_d }{(2\pi)^d} \nu p^2\frac{(d-1)}{ 4 (d+2)}  \frac{w}{\epsilon}\,\Bigg|_{d=4}= 
     - \frac{1}{8\pi^2} \nu p^2\frac{1}{ 8}  \frac{w}{\epsilon}. \label{B4}
\end{eqnarray}
Here $S_d =2\pi^{d/2}/\Gamma(d/2)$ is the area of the unit sphere in the $d$-dimensional space. 

By redefining charge $w/8\pi^2\rightarrow w$ and putting result (\ref{B4}) in Eq. (\ref{vv'}), one arrives at one-loop approximation for renormalization constant $Z_3$:
\begin{eqnarray}
    Z_3 = 1 - \frac{1}{ \epsilon} \frac{w}{8}. \label{z3}
\end{eqnarray}

To find $Z_1$ and $Z_2$ we have to calculate the second diagram in Eq. (\ref{hh'}):  
\begin{eqnarray}
    \hhstrokeSeconddetail\,= 
    \int_{k>m} \frac{d\omega}{2\pi}\frac{d^d {\bf k}}{(2\pi)^d} \frac{ D_{0} P_{ab} ({\bf k}) \, i p_a i (p-k)_b}{(\omega^2+\nu ^2_0 k^4)(-i\omega+\epsilon(p-k))}.
\label{hh' light 1}
\end{eqnarray}
By integrating over the frequency $w$ (using residues) and dropping the factor $k_b$ from the numerator, we arrive at the following expression: 
\begin{eqnarray}
    - \frac{D_{0}}{2 \nu_0} \int_{k>m} \frac{d^{d-1} {\bf k_\perp} d k_\parallel}{(2\pi)^d} \frac{ P_{ab} ({\bf k}) p_a p_b }{k^2 ((\nu_0 + \nu_{\parallel 0}) k^2_\parallel + (\nu_0 + \nu_{\perp 0}) k^2_\perp )}.
\label{hh' light 2}
\end{eqnarray}
The numerator $P_{ab} ({\bf k}) p_a p_b$ can be decomposed into the sum 
\begin{eqnarray}
\left(p^2 k^2 - (p_\parallel k_\parallel)^2 - (p_\perp k_\perp)^2 \right) /k^2,
\end{eqnarray}
so well-known relation for $d$-dimensional integral
\begin{eqnarray}
  \int d^d {\bf k} \frac{k_i k_j}{k^2} f(k) = \frac{\delta_{ij}}{d} \int d^d {\bf k} f(k), 
\end{eqnarray}
($f(k)$ here is an arbitrary function that depends only on $k$) allows us to transform integral (\ref{hh' light 2}) into the following one:
\begin{eqnarray}
         - \frac{D_{0}}{2 \nu_0} \int_{k>m} \frac{d^{d-1} {\bf k_\perp} d k_\parallel}{(2\pi)^d} \frac{ \left[  p^2 k^2 - p^2_\parallel k^2_\parallel - \frac{p^2_\perp k^2_\perp}{d-1} \right] }{k^4 ((\nu_0 + \nu_{\parallel 0}) k^2_\parallel + (\nu_0 + \nu_{\perp 0}) k^2_\perp )}  =\nonumber\\
   - \frac{D_{0}}{2 \nu_0} \int_{k>m} \frac{d^{d-1} {\bf k_\perp} d k_\parallel}{(2\pi)^d} \frac{ k^2_\parallel p^2_\perp + k^2_\perp \left(p^2 - \frac{p^2_\perp}{d-1} \right) }{k^4 ((\nu_0 + \nu_{\parallel 0}) k^2_\parallel + (\nu_0 + \nu_{\perp 0}) k^2_\perp )}.
    \label{hh' light 3}
\end{eqnarray}
Integration over $k_\parallel$ by residue produces 
\begin{eqnarray}
    \frac{- D_{0}}{16 \nu_0 \left(\sqrt{\nu_0 + \nu_{\parallel 0}} + \sqrt{\nu_0 + \nu_{\perp 0}} \right)^2} \int_{k>m} \frac{d^{d-1} {\bf k_\perp} }{(2\pi)^{d-1}} \frac{1}{ \left| k_\perp \right|^3}\cdot \nonumber\\
    \left[ p^2_\perp + \left( p^2 - \frac{p^2_\perp}{d-1} \right) \left( 1 + 2\sqrt{\frac{\nu_0 + \nu_{\parallel 0}}{\nu_0 + \nu_{\perp 0}}} \right) \right].
\label{hh' light 5}
\end{eqnarray}
Reducing integral over momenta to the dimensionless scalar integral as follows
\begin{eqnarray}
        \int_{k>m} \frac{d^d {\bf k}}{k^{d+\epsilon}} = \frac{S_d}{\epsilon} \frac{1}{m^\epsilon}, 
\end{eqnarray}
we arrive at the final result for (\ref{hh' light 1}):
\begin{eqnarray}
   \hhstrokeSecond &=& - \frac{1}{8 \pi^2}\cdot\frac{w\nu ^2}{ 2 \epsilon\left(\sqrt{\nu  + \nu_{\parallel}} + \sqrt{\nu  + \nu_{\perp}} \right)^2} \cdot \nonumber\\
    &&\left[ p^2_\parallel \left( 1 + 2\sqrt{\frac{\nu  + \nu_{\parallel}}{\nu  + \nu_{\perp }}} \right) +     \frac{p^2_\perp}{3} \left( 5 + 4\sqrt{\frac{\nu  + \nu_{\parallel }}{\nu  + \nu_{\perp }}} \right) \right] .
\label{hh' light 6}
\end{eqnarray}
Note that we put $d = 4$ and substituted all quantities with the first order of their decomposition into their renormalized counterparts (in case of MS scheme, this implies that renormalization constants were substituted with a unity). 

Let us rewrite this expression using several new notations; firstly, let $w$ again stand for $w/8\pi^2$. Secondly, charges $x_1$, $x_2$ should be used instead of ratios $\nu_{\parallel }/\nu $ and $\nu_{\perp }/\nu $. Finally, let us introduce the functions $f_1(x_1,x_2)$ and $f_2(x_1,x_2)$:
\begin{eqnarray}
   f_1(x_1,x_2) =\frac{1}{ 2 x_1\left(\sqrt{1 + x_1} + \sqrt{1 + x_{2}} \right)^2}\left( 1 + 2\sqrt{\frac{1 + x_1}{1 + x_{2}}} \right);
   \label{ff}   \\
   f_2(x_1,x_2) =\frac{1}{ 6 x_2 \left(\sqrt{1 + x_1} + \sqrt{1 + x_{2}} \right)^2} 
    \left( 5 + 4\sqrt{\frac{1 + x_{1}}{1 + x_{2}}} \right) .
    \label{fff}
\end{eqnarray}
Then we arrive at the following expression:
\begin{eqnarray}
   \hhstrokeSecond = - \frac{w}{ \epsilon} \, 
    \left[ \nu_{\parallel }p^2_\parallel f_1(x_1,x_2) +    \nu_{\perp } p^2_{\perp}f_2(x_1,x_2)  \right] .
    \label{fin}
\end{eqnarray}

Equations (\ref{hh'}), (\ref{z3}) combined with Eq. (\ref{fin}) give us one-loop results for renormalization constants $Z_1$ and $Z_2$:
\begin{eqnarray}
    Z_1 = 1-\frac{1}{\epsilon }\left[ g\frac{3}{16}+w\, f_1(x_1,x_2) \right], &  \\ 
    Z_2 = 1 - \frac{1}{ \epsilon} w\,f_2(x_1,x_2) . &  
\end{eqnarray}
From now on, we change the notation as $g/8\pi^2\rightarrow g$. 

Finally, consider the one-loop approximation for the  1-irreducible Green's function $\langle h' hh \rangle_{\rm 1-ir}$:
\begin{eqnarray}
\langle h'hh \rangle_{1-ir} = 
\raisebox{-1ex}[0cm][0cm]{\begin{tikzpicture}
\begin{feynman}
\vertex (a1) ;
					\vertex[below=0.4cm of a1] (a2);
				    \vertex[below left=0.4cm of a2] (a5);
				    \vertex[below right=0.4cm of a2] (a6);
					\diagram{
						(a1) --[fermionZerowidth'''] (a2),
						(a2) --[fermion] (a5),
						(a2) --[fermion](a6),
					};
				\end{feynman}
\end{tikzpicture}} +
\raisebox{-4ex}[0cm][0cm]{\begin{tikzpicture}
\begin{feynman}
\vertex (a1) ;
					\vertex[below=0.4cm of a1] (a2);
					\vertex[below right= 1.5cm of a2] (a3);
					\vertex[below left= 1.5cm of a2] (a4);
				    \vertex[below left=0.4cm of a4] (a5);
				    \vertex[below right=0.4cm of a3] (a6);
					\diagram{
						(a1) --[fermionZerowidth'''] (a2),
						(a2) --[fermion'] (a3),
						(a2) --[fermion'](a4),
						(a5) --[fermion] (a4),
						(a3) --[fermion](a4),
						(a3) --[fermion](a6),
					};
				\end{feynman}

\end{tikzpicture}} + 
\raisebox{-4ex}[0cm][0cm]{\begin{tikzpicture}
\begin{feynman}
\vertex (a1) ;
					\vertex[below=0.4cm of a1] (a2);
					\vertex[below right= 1.5cm of a2] (a3);
					\vertex[below left= 1.5cm of a2] (a4);
				    \vertex[below left=0.4cm of a4] (a5);
				    \vertex[below right=0.4cm of a3] (a6);
					\diagram{
						(a1) --[fermionZerowidth'''] (a2),
						(a2) --[fermion] (a3),
						(a2) --[fermion'](a4),
						(a4) --[fermion] (a5),
						(a4) --[fermion'](a3),
						(a3) --[fermion](a6),
					};
				\end{feynman}

\end{tikzpicture}} + 
\raisebox{-4ex}[0cm][0cm]{\begin{tikzpicture}

\begin{feynman}
	\vertex (a1) ;
					\vertex[below=0.4cm of a1] (a2);
					\vertex[below right= 1.5cm of a2] (a3);
					\vertex[below left= 1.5cm of a2] (a4);
				    \vertex[below left=0.4cm of a4] (a5);
				    \vertex[below right=0.4cm of a3] (a6);
					\diagram{
						(a1) --[fermionZerowidth'''] (a2),
						(a2) --[fermion] (a3),
						(a2) --[fermion'](a4),
						(a4) --[fermion] (a5),
						(a4) --[fermion'](a3),
						(a3) --[fermion](a6),
					};
				\end{feynman}

\end{tikzpicture}} \nonumber \\ 
+\raisebox{-4ex}[1.7cm][0cm]{\begin{tikzpicture}
\begin{feynman}
							\vertex (a1) ;
					\vertex[below=0.4cm of a1] (a2);
					\vertex[below right= 1.5cm of a2] (a3);
					\vertex[below left= 1.5cm of a2] (a4);
				    \vertex[below left=0.4cm of a4] (a5);
				    \vertex[below right=0.4cm of a3] (a6);
					\diagram{
						(a1) --[fermionZerowidth'''] (a2),
						(a2) --[fermion'] (a3),
						(a2) --[fermion'](a4),
						(a5) --[fermion] (a4),
						(a3) --[Boson](a4),
						(a3) --[fermion](a6),
					};
				\end{feynman}

\end{tikzpicture}} +  
\raisebox{-4ex}[1.7cm][0cm]{\begin{tikzpicture}
\begin{feynman}
							\vertex (a1) ;
					\vertex[below=0.4cm of a1] (a2);
					\vertex[below right= 1.5cm of a2] (a3);
					\vertex[below left= 1.5cm of a2] (a4);
				    \vertex[below left=0.4cm of a4] (a5);
				    \vertex[below right=0.4cm of a3] (a6);
					\diagram{
						(a1) --[fermionZerowidth'''] (a2),
						(a2) --[fermion'] (a3),
						(a2) --[Boson](a4),
						(a5) --[fermion] (a4),
						(a3) --[fermion'](a4),
						(a3) --[fermion](a6),
					};
				\end{feynman}

\end{tikzpicture}} + 
\raisebox{-4ex}[1.7cm][0cm]{\begin{tikzpicture}

\begin{feynman}
							\vertex (a1) ;
					\vertex[below=0.4cm of a1] (a2);
					\vertex[below right= 1.5cm of a2] (a3);
					\vertex[below left= 1.5cm of a2] (a4);
				    \vertex[below left=0.4cm of a4] (a5);
				    \vertex[below right=0.4cm of a3] (a6);
					\diagram{
						(a1) --[fermionZerowidth'''] (a2),
						(a2) --[Boson] (a3),
						(a2) --[fermion'](a4),
						(a4) --[fermion] (a5),
						(a4) --[fermion'](a3),
						(a3) --[fermion](a6),
					};
				\end{feynman}
\end{tikzpicture}}\raisebox{-2ex}[1.5cm][0cm]{.}
\label{h'hhir}
\end{eqnarray}
\\

The UV divergences in the first three diagrams cancel each other out due to Galilean symmetry of the Hwa-Kardar equation (\ref{HK}): $h(t,{\bf x}) \to h(t,{\bf x}+u{\bf n}\,t)-u{\bf n}$, where $u=$const. If one considered possible counterterms to the action functional that corresponds to the ``pure'' Hwa-Kardar equation without turbulent mixing, this symmetry would forbid the counterterm $h'\partial_{\parallel} h^2$. In practice, this fact translates to cancellation of the UV divergences.

The divergent part of the three remaining diagrams (the diagrams in the second line of Eq. (\ref{h'hhir})) vanish due to incompressibility of velocity field ${\bf v}$. Indeed, without loss of the generality the routing of the external momenta can be chosen such that corresponding integrands would appear to be proportional to $P_{ab}({\bf k})\,k_b=0$, where ${\bf k}$ is the momentum flowing through a propagator $\langle{\bf v\,v}\rangle_0$, so each of the integrals would vanish as a whole.

As a result, the renormalization constant $Z_4$ appears trivial in the one-loop approximation: $Z_4=1$. It is tempting to assume that the symmetry and the  incompressibility would keep this relation exact in all orders of perturbation theory. However, while the symmetry guarantees that all diagrams without the velocity field ${\bf v}$ cancel each other in all orders, the same cannot be said about the incompressibility. Thus, the  found value of $Z_4$ cannot be claimed to be exact. 

The one-loop results for renormalization constants are:
\begin{eqnarray}
Z_1 = 1-\frac{1}{\epsilon }\left[ g\frac{3}{16}+w\, f_1(x_1,x_2) \right],  \nonumber\\ 
    Z_2 = 1 - \frac{1}{ \epsilon} w\,f_2(x_1,x_2), \nonumber \\  
Z_3 = 1 - \frac{1}{ \epsilon} \frac{w}{8}, \nonumber\\
Z_4=1,
\label{ZZZZ}
\end{eqnarray}
with the functions $f_{1,2}$ defined in (\ref{ff}) and (\ref{fff}) and the higher-order corrections involving higher powers in $w$ and $g$.

\section{\label{sec:levelA}RG functions and attractors of the RG equations}

Let us define renormalization functions (anomalous dimensions $\gamma$ and $\beta$  functions):
\begin{eqnarray}
\gamma_{Q} = \widetilde{\cal D}_{\mu}\ln Z_{Q},\nonumber\\
\beta_r = \widetilde{\cal D}_{\mu}r.
\label{XXX}
\end{eqnarray}
Here $Q$ is a given quantity with renormalization constant $Z_{Q}$ and $r$ is a coupling constant. Differential operator $\widetilde{\cal D}_{\mu}$
\begin{equation}
\widetilde{\cal D}_{\mu} = \mu\partial_{\mu}|_{\{g_0, w_0, x_{10}, x_{20}, \nu_0\}}
\end{equation}
emerges from the relation $\widetilde{\cal D}_{\mu}F=0$ for a physical quantity $F$ that encapsulates the fact that $F$ cannot depend on renormalization mass $\mu$ (which is not an observable).

Anomalous dimensions for the model (\ref{action_ve}) are
\begin{eqnarray}
    \gamma_{h} = -\gamma_{h'}=\gamma_{4}, \quad \gamma_{{\bf v}}=\gamma_{{\bf v'}}=0,
    \nonumber \\ \gamma_g = -3 \gamma_{1}/2-3\gamma_{2}/2+2\gamma_4, \quad \gamma_w = -3\gamma_{3}, \quad \gamma_{x_1} = \gamma_{1}-\gamma_{3}, \quad \gamma_{x_2} = \gamma_{2}-\gamma_{3}, \nonumber \\
    \gamma_{\nu_{\parallel}}=\gamma_1, \quad \gamma_{\nu_{\perp}}=\gamma_2, \quad  \gamma_{\nu}=\gamma_3,
\label{gamma}
\end{eqnarray}
while one-loop approximation for $\gamma_1$--$\gamma_4$ is
\begin{eqnarray}
    \gamma_1 =  g\frac{3}{16}+w\,f_1(x_1,x_2), \quad
    \gamma_2 = w \,f_2(x_1,x_2), \quad
    \gamma_3 =  \frac{w}{8}, \quad \gamma_{4} = 0. 
    \label{op}
\end{eqnarray}
The $\beta$ functions for the coupling constants $g$, $w$, $x_1$, and $x_2$ read
\begin{eqnarray}
\beta_g = -g \left[\epsilon +\gamma_{g} \right], \quad \beta_w = -w \left[\epsilon + \gamma_w \right], \quad \beta_{x_1} = -x_1 \gamma_{x_1}, \quad \beta_{x_2} = -x_2 \gamma_{x_2}.
    \label{bthroughg}
\end{eqnarray}

In the one-loop approximation, the $\beta$  functions have the forms: 
\begin{eqnarray}
    \beta_{g} =& -g\left[ \epsilon - \frac{9}{32}g - \frac{3}{2}w\,\left(f_1(x_1,x_2)+f_2(x_1,x_2)\right)\right]
    \label{b1}
\end{eqnarray}
\begin{equation}
    \beta_{w} = -w\left[ \epsilon - \frac{3}{8} w \right],
    \label{b2}
\end{equation}
\begin{equation}
    \beta_{x_{1}} = -x_{1}\left[ \frac{3}{16}g - \frac{1}{8}w + w \,f_1(x_1,x_2) \right],
    \label{b3}
 \end{equation}
\begin{equation}
    \beta_{x_{2}} = -x_{2}\left[ - \frac{1}{8}w + w \,f_2(x_1,x_2) \right].
    \label{b4}
\end{equation}

The functions (\ref{bthroughg}) satisfy the exact identity
\begin{equation}
    \beta_{g} = -g \left[ \frac{3}{2} \frac{\beta_{x_{1}}}{x_{1}} + \frac{3}{2} \frac{\beta_{x_{2}}}{x_{2}} - \frac{\beta_{w}}{w} + 2\gamma_{4}\right]
    \label{lineardependence}
\end{equation}
that follows from the definitions (\ref{XXX}) and the relations (\ref{gamma}).
 
In the one-loop approximation, one has $\gamma_{4}=0$, so that the relation 
(\ref{lineardependence}) becomes a linear dependence between the $\beta$  functions. As we will see below, this makes the most interesting attractor of the RG equations to be a curve of fixed points rather than a single point. 

Universality classes of the IR behaviour are defined by the critical exponents and the coordinates of the RG equation fixed points. The latter 
are provided  by the zeroes of the $\beta$ functions system: $\beta(g^*)=0$; see, {\it e.g.}~\cite{Book3} (Section~1.42). Furthermore, to correspond to IR range behavior, a fixed point must be IR attractive, {\it i.e.} the eigenvalues $\lambda_i$ of matrix $\Omega_{rn}=\{\partial\beta_{g_r}/\partial g_n\}|_{g^*}$ must have strictly positive real parts; see, {\it e.g.}~\cite{Book3} (Section~1.42).

Taking into account that charges $x_1$, $x_2$ are positive and charge $g$ is non-negative, we find the following fixed points:

FP1 -- a two-dimensional surface of Gaussian (trivial) fixed points:
\begin{eqnarray}
    g^{*} = 0; \quad w^{*} = 0; \quad x_{1}^{*}\neq 0; \quad x_{2}^{*} \neq 0;\nonumber\\
    \lambda_i=\{0,0,-\epsilon,-\epsilon\}.
    \label{FP1}
\end{eqnarray}

\begin{figure}[h]
    \centering
    \includegraphics[width=0.8\textwidth]{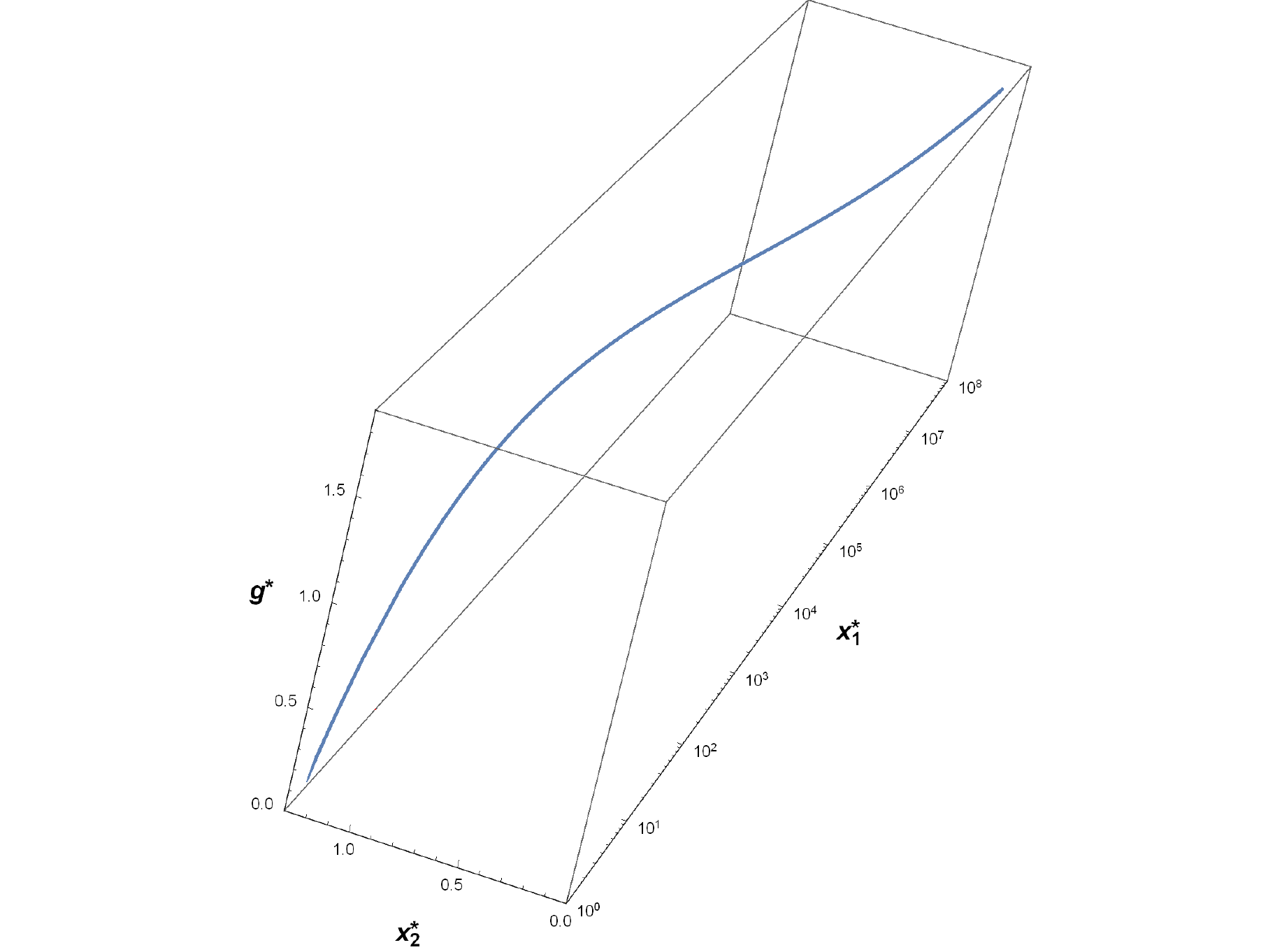}
    \caption{Curve of fixed points FP2 in three-dimensional space of coordinates $\{g^*(x^{*}_{2}),\  x^{*}_{1}(x^{*}_{2}),\ x^{*}_{2}\}$ (we recall that $w^*=8\epsilon/3$).}
    \label{curve}
\end{figure}

FP2 -- a curve of fixed points, parametrized by one of the coordinates ({\it e.g.} $\{g^*(x^{*}_{2}),\ w^*,\ x^{*}_{1}(x^{*}_{2}),\ x^{*}_{2}\}$) and determined by the following expressions:

\begin{eqnarray}
    g^{*} = \frac{16\epsilon}{9}\left(1-8f_1(x_1^{*},x_2^{*})\right);\quad
    w^{*} = \frac{8}{3}\epsilon; \quad
   f_2(x_1^{*},x_2^{*}) =\frac{1}{8};\nonumber\\
    \lambda_i=\{0,\epsilon,\lambda_3,\lambda_4\}.
    \label{FP2}
\end{eqnarray}
The curve is ``semi-infinite'' as two charges span finite segments of values while charge $x_1$ runs through $\left(+\infty, (\sqrt{13} - 1)/2\right]$ (see Fig.\ref{curve}).

\begin{figure}[h]
    \centering
    \includegraphics[width=0.6\textwidth]{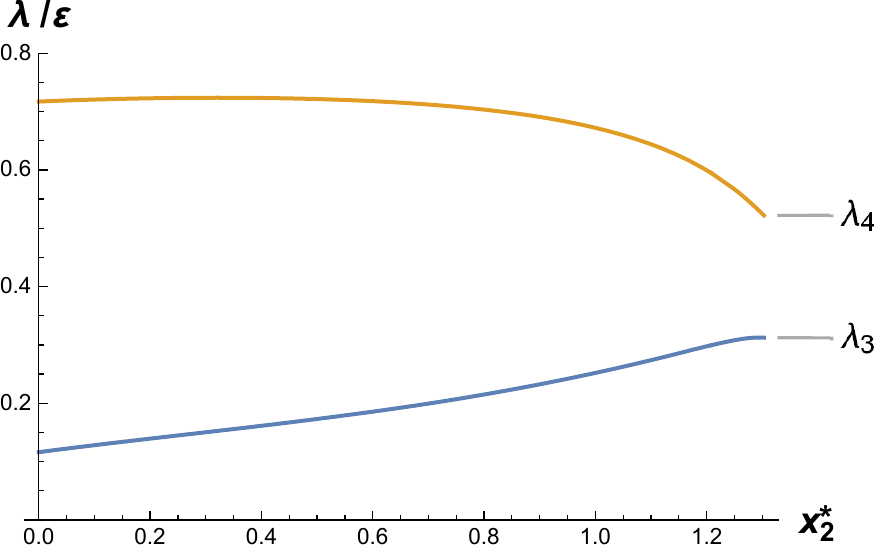}
    \caption{Eigenvalues $\lambda_3$ and $\lambda_4$ parameterized by $x_2^*$.}
    \label{check3}
\end{figure}

The eigenvalues $\lambda_3$ and $\lambda_4$ change along the curve FP2 and can also be parametrized, see Fig.\ref{check3}. Both the eigenvalues are non-negative for the relevant values of $x_2^*$ and positive values of $\epsilon$.

The curve FP2 includes a special point where the HK nonlinearity becomes IR irrelevant:

FP2a:  
\begin{eqnarray}
    g^{*} = 0; \quad w^{*} = \frac{8}{3}\epsilon; \quad x_{1}^{*}=x_{2}^{*} = \frac{1}{2}(\sqrt{13} - 1);\nonumber\\
    \lambda_i=\left\{0,\epsilon, \frac{47 + \sqrt{13}}{162} \epsilon, \frac{13 - \sqrt{13}}{18} \epsilon \right\}.
    \label{FP2a}
\end{eqnarray}

This endpoint corresponds to the marginal case where the nonlinearity of the original HK equation becomes completely irrelevant making the dynamics isotropic in the leading order of the IR asymptotic behaviour.

The existence of the {\it curve} of fixed points is a consequence of the exact equation (\ref{lineardependence}) which turns to a linear relation between the $\beta$  functions in the one-loop approximation, where $\gamma_4=0$, see the last relation in (\ref{op}).  

However, there are no obvious reasons for the relation $\gamma_4=0$ in (\ref{op}) to hold in all orders of perturbation theory. 
Thus, at the one-loop level, it is impossible to determine whether we are dealing with a curve or a single point ``in disguise'', {\it i.e.} whether the curve FP2 would be reduced to a single point, if one were to take into account all higher order corrections.

Nevertheless, it is indeed peculiar that the point FP2a lies on the curve FP2. Fixed points similar to FP2a arise in most problems where stochastic system is affected by environment motion; it is natural to expect appearance of such a point. However, it now shares stability region with all the other points on the curve FP2.

The entire curve is IR stable simultaneously (see Fig. \ref{check3}) which means that points with different coordinates and different values of eigenvalues are IR attractive for the same values of parameter $\epsilon$. It is tempting to explain away this odd occurrence as an artefact of one-loop approximation and to expect the curve to shrink to its endpoint FP2a in higher order approximations.

As we will see in the next Section, the critical dimensions of the fields are the same for all points on the curve FP2, 
while the correction exponents, determined by the eigenvalues, vary along the line. Regime of critical behavior related to the point FP2a (in which the isotropic environment dominates the dynamics) thus ``extends'' to the entire curve despite the fact that other points on the curve has non-zero coordinate $g^*$, {\it i.e.} the non-linearity of the Hwa-Kardar equation is relevant.

\begin{figure}
    \centering
    \subfigure[]{\includegraphics[width=0.49\textwidth]{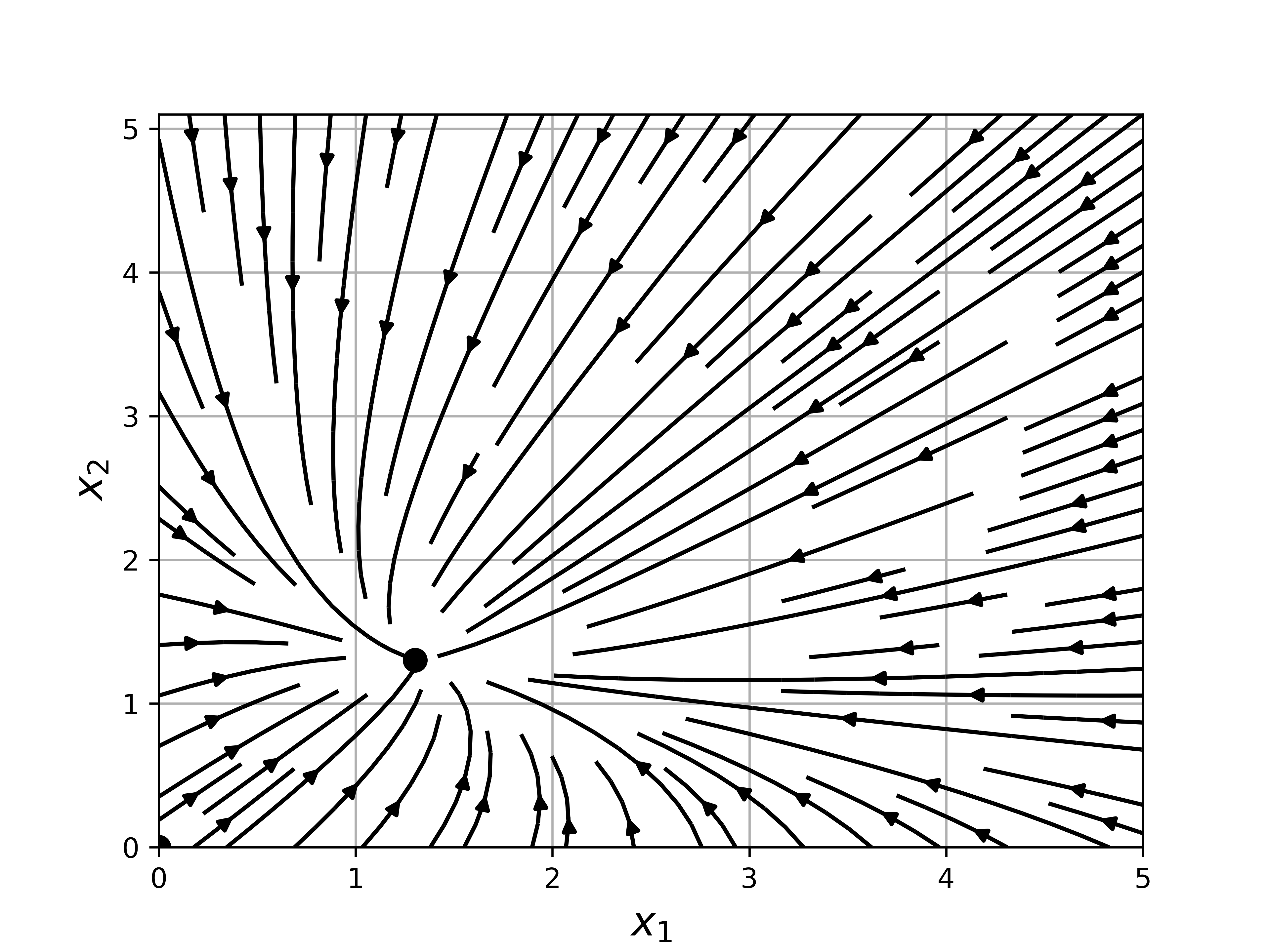}} 
    \subfigure[]{\includegraphics[width=0.49\textwidth]{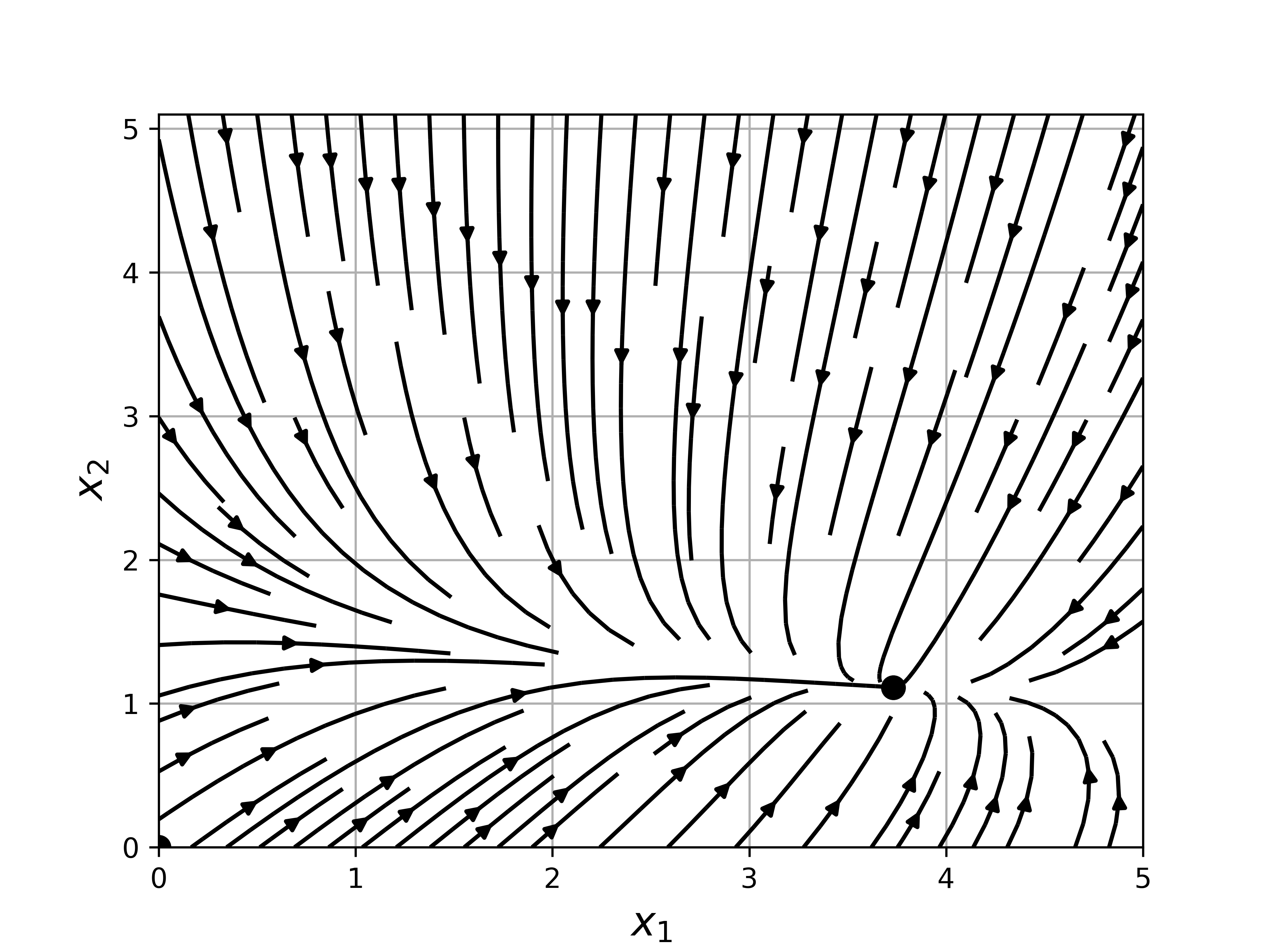}} 
    \caption{RG flows for positive $\epsilon$; arrows indicate movement towards IR limit; (a) RG flow for $g^*=0$; the dot is the fixed point from FP2 with $g^*=0$. (b) RG flow for $g^*=1,2$; the dot is the fixed point from FP2 with $g^*=1,2$. }
    \label{fig:test}
\end{figure}

If the curve FP2 exists in all orders of perturbation theory, then coinciding stability regions imply a loss of universality, {\it i.e.} that critical behavior of the system depends on local parameters. Direct numerical simulations of the RG flows\footnote{RG flow is defined by the system of ordinary differential equations ${\cal D}_{s} \overline{g}_i = \beta_i(\overline{g}_j)$ for the invariant (``running'') coupling constants $\overline{g}_i(s,g)$, where $g=\{g_i\}$ is the full set of coupling constants and $s=k/\mu$ features some momentum variable $k$; see, {\it e.g.}  Section~1.42~in~\cite{Book3}.} 
(based on one-loop approximation for $\beta$  functions) for positive $\epsilon$ show that each point on the curve is indeed IR-attractive, see Fig. \ref{fig:test}.

We also considered marginal values of $x_1^*$ and $x_2^*$, where $x_1^*, x_2^*=0$ and $x_1^*, x_2^*\rightarrow\infty$. To do that one should pass to another set of charges, ({\it e.g.} $g$, $w$, $\alpha_1=x_1^{-1}$, $\alpha_2=x_2^{-1}$), and investigate possible fixed points. However, all of these points turned out to have nearly empty stability regions defined, {\it e.g.} by contradictory inequalities like $\epsilon>0$ and $\epsilon<0$. For example, the fixed point related to the regime where only nonlinearity of HK equation is relevant with coordinates $g^*=32\epsilon/9$, $w^*=0$, $\alpha_1^*=0$, arbitrary $\alpha_2^*$, has eigenvalues $\{0,-\epsilon,2\epsilon/3,\epsilon\}.$ Such ``saddle'' points can be useful  for general analysis of RG flows but they do not correspond to possible critical behavior in the same way IR attractive fixed points do. Thus, we do not investigate them further.

\section{Critical dimensions} \label{sec:CD}

Critical behaviour of the Green's functions is described by critical dimensions (related to critical exponents) or powers in the scaling law in the IR range. Critical dimensions are defined for IR relevant parameters, {\it i.e.} for the parameters that are dilated in critical scaling as opposed to IR irrelevant quantities (in our case, that includes renormalization mass $\mu$ and all four coupling constants) that stay fixed. Expression
\begin{equation}
        \Delta_{F} = d^k_{F} + \Delta_\omega d^{\omega}_{F} + \gamma_{F}^*,
    \label{critical dimension}
\end{equation}
gives a critical dimension $\Delta_{F}$ of a field $F$ (see, {\it e.g.}~Section~2.1 in~\cite{RedBook}, Section~3.1 in~\cite{UFN} and Sections~5.16 and 6.7 in~\cite{Book3}). Here $\gamma_{F}^*$ is anomalous dimension of the field taken at some IR attractive fixed point. Frequency critical dimension $\Delta_\omega$ is given by the expression
\begin{equation}
\Delta_\omega = 2 - \gamma_{\nu}^*.
\label{Domega}
\end{equation}
For the fixed points FP1, all the critical dimensions are found exactly:
\begin{equation}
\Delta_h=\Delta_{{\bf v}}=1, \quad 
\Delta_{h'}=\Delta_{{\bf v'}}=d-1, \quad \Delta_{\omega} =2.
\label{DimFP1}       
\end{equation}
For the curve of fixed points FP2 (including the endpoint FP2a), the critical dimensions
\begin{equation}
\Delta_h=\Delta_{{\bf v}}=1-\epsilon/3, \quad \Delta_{h'}=\Delta_{{\bf v'}}=d-1+\epsilon/3, \quad \Delta_{\omega} =2-\epsilon/3
\label{DimFP2}       
\end{equation}
are also known perturbatively exactly in the sense that they have no higher-order corrections in $\varepsilon$ (see discussion in the next section). As it should be, the dimensions $\Delta_{{\bf v}}$, $\Delta_{{\bf v'}}$ and $\Delta_{\omega}$ for FP2a coincide with those derived in \cite{FNS}.

The fixed point $g^*=32\epsilon/9$, $w^*=0$, $\alpha_1^*=0$, arbitrary $\alpha_2^*$ mentioned in Sec. \ref{sec:levelA} allows for a kind of dimensional transmutation,
in the sense that the ratio of the diffusivity coefficients $\nu_{\parallel}$ and $\nu_{\perp}$ acquires nontrivial canonical and critical dimensions. A new canonical symmetry arises that allows one to introduce two independent momentum canonical dimensions $d^{\parallel}$ and $d^{\perp}$ for the two subspaces. 
The scaling behavior corresponding to that fixed point is the one where the coordinates $x_{\parallel}$ and $x_{\perp}$ are diluted in the different way resulting in nontrivial critical dimension $\Delta_{\parallel}$. However, in our model, that fixed point is never IR attractive and this interesting scaling regime cannot realize for general initial conditions. 
A thorough discussion of this dimensional transmutation can be found
in \cite{WeU,WeU2}.

\section{Conclusion}\label{sec:Conc}

We studied a semi-phenomenological, strongly anisotropic continuous model of a system with self-organized criticality (\ref{HK})--(\ref{forceC})
proposed in \cite{HK} and subjected to isotropic randomly moving environment modelled by the Navier-Stokes stochastic differential equation for an incompressible viscous fluid (\ref{NS})--(\ref{forceD}) inspired by \cite{FNS}.

We constructed the corresponding  field theoretic model (\ref{action_ve}) and established its renormalizability. The renormalization constants, the RG functions and the coordinates of the IR attractors were calculated in the leading one-loop approximation. The RG analysis showed that the IR behaviour of the model is governed by the IR attractor, referred to as FP2 in Section~\ref{sec:levelA}, that is a curve of fixed points rather than a single point.
Moreover,
the critical dimensions of the basic quantities (the fields and the frequency) remain the same along the curve. Thus, it is natural and tempting to interpret the whole curve as a single universality class.

In this respect, our situation differs from those where the attractors of the RG flows were given by lines \cite{AKa,AntK,Basu} or surfaces \cite{Kohmoto}-\cite{ABK}
with the critical exponents continuously varying along the attractive parts of the manifolds. In our case, only the correction exponents (the eigenvalues of the stability matrix $\Omega$) vary along the curve.

It should be stressed that the appearance of the curve of fixed points is a consequence of the linear relation  between the $\beta$ functions that results from (\ref{lineardependence}) for $\gamma_4=0$. The latter expression follows form the one-loop result $Z_4=1$ in (\ref{ZZZZ}), which, in its turn, reflects the absence of a counterterm $h'\partial_{\parallel}h^2$.
However, this counterterm is not forbidden by the real index of divergence nor some evident symmetry, so we can see no clear reason to expect that the relations $Z_4=1$ and  $\gamma_4=0$ are in fact exact and hold in all orders of perturbation theory.

The two-loop calculations are expected to be rather difficult for our complicated model. However, even without such a calculation it is clear that there are two main possibilities.

First option: for some reasons, like a certain obscure hidden symmetry, the last relation $Z_4=1$ in (\ref{ZZZZ}) and its consequence $\gamma_4=0$ in (\ref{op}) are exact identities. Then the IR attractor is indeed a curve. The critical dimensions of the fields $h'$ and $h$ remain the same along the curve and are given exactly by  (\ref{DimFP2}). The exactness of those expressions follow from (\ref{critical dimension}) and (\ref{Domega}). Indeed, the anomalous dimensions of the fields $h$ and $h'$ are equal (up to the sign) with the {\it vanishing} anomalous dimension $\gamma_{4}$. From the second equality in (\ref{bthroughg}) it follows that $\gamma_{w}^{*} = - \varepsilon$ along the curve, while the  exact relation $\gamma_{\nu} = - \gamma_{w}/3$ 
delivers the exact value of the $\gamma_{\nu}^{*} = \varepsilon/3$ 
for the whole curve FP2.
As a result, in the option consireded,  only the correction exponents (eigenvalues of the matrix  $\Omega$) which are, in principle, observable quantities, vary along the curve of fixed points.

Second option: the relation $\gamma_4=0$ in (\ref{op}) is an artefact of the one-loop approximation and is violated by the higher-order corrections.  
Then the IR attractive curve will shrink to a single point satisfying the equation $\gamma_{4}^{*} = 0$ following from (\ref{lineardependence}). From physics reasons, it is clear that this point corresponds to the regime when the HK non-linearity becomes irrelevant and the expressions (\ref{DimFP2}) for the critical dimensions are in fact exact. This is the point referred to as FP2a in Section~\ref{sec:levelA}.

In the future, it would be interesting to consider other types of the velocity ensemble, especially the power-like correlation function of the stirring force that describes turbulent motion \cite{RedBook,UFN} or its combination with the simple stirring (\ref{forceD1}) due to \cite{FNS}.
It is also desirable to go beyond the leading one-loop approximation. 
This work is already in progress.

\section*{Acknowledgments}

The authors are indebted to Abhik~Basu for valuable discussion and for bringing our attention to the paper~\cite{Basu}. One of the authors (N.V.A.)
is indebted to M.~I.~Vyazovsky for enlightening discussions.

The work was supported by the Foundation for the Advancement of Theoretical Physics and Mathematics ``BASIS'' (P.~I.~K., project number 22-1-3-33-1) and by Ministry of Science and Higher Education of the Russian Federation, agreement \textnumero  075–15–2022–287 (P.~I.~K. and A.~Yu.~L.).


\section*{References}
\begin{thebibliography}{199}

\bibitem{Bak} Bak P 1999 {\it How Nature Works: the Science of Self-Organized Criticality} (Copernicus)

\bibitem{Amit} Amit D J 1984 {\it Field Theory, Renormalization Group, and Critical Phenomena 2nd edition}  (World Scientific, Singapore)

\bibitem{Zinn} Zinn-Justin J 1989 {\it Quantum Field Theory and Critical Phenomena} (Clarendon Press: Oxford)

\bibitem{Book3} Vasiliev A N 2004 {\it The Field
Theoretic Renormalization Group in Critical Behavior Theory and Stochastic Dynamics} (Chapman \& Hall/CRC, Boca Raton) 

\bibitem{Hinrichsen} Hinrichsen H 2000 {\it Non-Equilibrium Critical Phenomena and Phase Transitions into Absorbing States} Advances in Physics {\bf 49} 815-958
 
\bibitem{Henkel} Henkel M, Hinrichsen H, L\"ubeck S and  Pleimling M 2008 {\it Non-Equilibrium Phase Transitions} vol~1 (Springer, Dordrecht)


\bibitem{BTW} Bak P, Tang C and Wiesenfeld K 1987 {\it Self-Organized Criticality: An Explanation of the $1/f$ noise} Phys. Rev. Lett. {\bf 59} 381

\bibitem{BTW1}
Tang C and Bak P 1988 {\it Critical exponents and scaling relations for self-organized critical phenomena} Phys. Rev. Lett. {\bf 60} 2347

\bibitem{BTW2}
Bak P and Sneppen K 1993 {\it Punctuated equilibrium and criticality in a simple model of evolution} Phys. Rev. Lett. {\bf 71} 4083

\bibitem{Bak1}
Jensen H J 1998 {\it Self-Organized Criticality: Emergent Complex behavior in Physical and Biological Systems} (Cambridge University Press, Cambridge)

\bibitem{Bak2} Turcotte D L 1999 {\it Self-organized criticality} Rep. Prog. Phys. {\bf 62} 1377

\bibitem{Bak3} Pruessner G 2012 {\it Self-Organized Criticality: Theory, Models and Characterisation} (Cambridge University Press)

\bibitem{Col0} Watkins N W, Pruessner G, Chapman S C, Crosby N B and Jensen H J 2016 {\it 25 years of self-organized criticality: Concepts and controversies} Space Sci. Rev. {\bf 198}, 3

\bibitem{Col1} Mu\~{n}oz M A 2018 {\it Colloquium: Criticality and dynamical scaling in living systems} Rev. Mod. Phys. {\bf 90} 031001

\bibitem{Col2} Markovic D and Gros C 2014 {\it Power laws and self-organized criticality in theory and nature} Phys. Rep. {\bf 536} 41

\bibitem{Col3} Aschwanden M J 2013 {\it Self-Organized Criticality Systems} (Open Academic Press, Berlin, Warsaw)

\bibitem{bio1} Ellis G F R and Kopel J 2019 {\it The dynamical emergence of biology from physics: branching causation via biomolecules} Front. Physiol. {\bf 9} 1966

\bibitem{bio2} Mora T and Bialek W 2011 {\it Are biological systems poised at criticality?} J. Stat. Phys. {\bf 144} 268

\bibitem{neu1} Hesse J and Gross T 2014 {\it Self-organized criticality as a fundamental property of neural systems} Front. Syst. Neurosci. {\bf 8} 166

\bibitem{neu2} Pasquale V, Massobrio P, Bologna L L,  Chiappalone M and Martinoia S 2008 {\it Self-organization and neuronal avalanches in networks of dissociated cortical neurons} Neuroscience {\bf 153} 1354

\bibitem{neu3} Orlandi J G, Soriano J, Alvarez-Lacalle E,  Teller S and Casademunt J 2013 {\it Noise focusing and the emergence of coherent activity in neuronal cultures} Nat. Phys. {\bf 9} 582

\bibitem{neu4} Timme N M, Marshall N J, Bennett N, Ripp M, Lautzenhiser E and Beggs J M 2016 {\it Criticality maximizes complexity in neural tissue} Front. Physiol. {\bf 7} 425

\bibitem{neu5} Kossio F Y K, Goedeke S, van den Akker B, Ibarz B, Memmesheimer R-M 2018 {\it Growing Critical: Self-Organized Criticality in a Developing Neural System} Phys. Rev. Lett. {\bf 121} 058301

\bibitem{neu6} Levina A, Herrmann J M and Geisel T 2007 {\it Dynamical synapses causing self-organized criticality in neural networks} Nat. Phys. {\bf 3} 857

\bibitem{net1} Tadi\'{c} B, Mitrovi\'{c} Dankulov M and Melnik R 2017 {\it Self-organised criticality and emergent hyperbolic networks: blueprint for complexity in social dynamics} Phys. Rev. E {\bf 96} 032307

\bibitem{net2}
Tadi\'{c} B 2019 {\it Self-organised criticality and emergent hyperbolic networks: blueprint for complexity in social dynamics} European Journal of Physics {\bf 40} 024002

\bibitem{net3} Tadi\'{c} B, Gligorijevic V, Mitrovic\'{c} M and Suvakov M 2013 {\it Co-evolutionary mechanisms of emotional bursts in online social dynamics and networks} Entropy {\bf 15} 5084

\bibitem{net4} Suvakov M and Tadi\'{c} B 2014 {\it Collective emotion dynamics in chats with agents, moderators and Bots} Condens. Matter Phys. {\bf 17} 33801

\bibitem{net5}
Holovatch Yu, Mrygold O, Szell M and Thurner S 2017 {\it Math Meets Myths: Quantitative Approaches to Ancient Narratives} ed R Kenna (Springer Int. Publishing)
 
\bibitem{net6} Kou G, Zhao Y, Peng Y and Shi Y 2012 {\it Multi-level opinion dynamics under bounded confidence}
PLoS One {\bf 7} e43507

\bibitem{crop} Torres-Rojo J M and Bahena-Gonz\'{a}lez R 2018 {\it Scale invariant behavior of cropping area losses}
Agricultural Systems {\bf 165} 33

\bibitem{autism} Tonello L, Giacobbi L, Pettenon A, Scuotto A, Cocchi M, Gabrielli F and Cappello G 2018 {\it Crisis Behavior in Autism Spectrum Disorders: A Self-Organized Criticality Approach} Complexity 5128157

\bibitem{Onuki2}
Onuki A and Kawasaki K 1980 {\it Critical phenomena of classical fluids under flow. I: Mean field approximation} Progr. Theor. Phys. {\bf 63} 122

\bibitem{Nelson}
Aronowitz A and Nelson D R 1984 {\it Turbulence in phase-separating binary mixtures} Phys. Rev. A {\bf 29} 2012

\bibitem{Satten} 
Satten G and Ronis D 1986 {\it Critical phenomena in randomly stirred fluids: Correlation functions, equation of motion, and crossover behavior} Phys. Rev. A {\bf 33} 3415

\bibitem{Nandy}
Nandy M K and Bhattacharjee J K 1998 {\it Renormalization-group analysis for the infrared properties of a randomly stirred binary fluid} J. Phys. A Math. Gen. {\bf 31} 2621

\bibitem{AHH} Antonov N V, Hnatich M and Honkonen J 2006 {\it Effects of mixing and stirring on the critical behavior} J. Phys. A: Math. Gen. {\bf 39} 7867


\bibitem{Parisi} Benzi R, Parisi G, Sutera A and Vulpiani A 1983 {\it A theory of stochastic resonance in climate changes} SIAM J. Appl. Math. {\bf 43} 565

\bibitem{Mitch} Feigenbaum M J, Procaccia I and Davidovich B 2001 {\it Dynamics of Finger Formation in Laplacian Growth Without Surface Tension} J. Stat. Phys. {\bf 103} 973–1007

\bibitem{Parisi1}
Giorgio Parisi–Facts–2021. NobelPrize.org. Nobel Prize Outreach AB 2021. 
Mon. 11 Oct. 2021. 
Available online: https://www.nobelprize.org/prizes/physics/2021/parisi/facts/ 
(accessed on 24 November 2022   ).

\bibitem{HK} Hwa T and Kardar M 1989 {\it Dissipative transport in open systems: An investigation of self-organized criticality} Phys. Rev. Lett. {\bf 62}(16) 1813

\bibitem{HK1} Hwa T and Kardar M 1992 {\it Avalanches, hydrodynamics and great events in models of sandpiles} Phys. Rev. A {\bf 45} 7002

\bibitem{FNS} Forster D, Nelson D R and Stephen M J 1977 {\it Large-distance and long-time properties of a randomly stirred fluid} Phys. Rev. A {\bf 16} 732

\bibitem{WeU} Antonov N V, Gulitskiy N M, Kakin P I and Kochnev G E 2020 {\it Effects of turbulent environment on self-organized critical behavior: isotropy vs. Anisotropy} Universe {\bf 6} 145

\bibitem{WeU2} Antonov N V, Gulitskiy N M, Kakin P I and  Semeikin M N 2022 {\it Dimensional transmutation and nonconventional scaling behaviour in a model of self-organized criticality}

\bibitem{FGV} 
Falkovich G, Gaw\c{e}dzki K and Vergassola M 2001 {\it Particles and fields in fluid turbulence} Rev. Mod. Phys. {\bf 73} 913

\bibitem{AKL} 
Antonov N V, Kakin P I and Lebedev N M 2019 {\it The Kardar–Parisi–Zhang model of a random kinetic growth: effects of a randomly moving medium} J. Phys. A: Math. Theor. {\bf 52} 505002

\bibitem{RedBook} 
Adzhemyan L T, Antonov N V and Vasil’ev A N 1999 {\it The Field Theoretic Renormalization Group in Fully Developed Turbulence} (Gordon \& Breach: London, UK)

\bibitem{UFN}
Adzhemyan L T, Antonov N V and Vasil’ev A N 1996 {\it Quantum field renormalization group in the theory of fully developed turbulence} Phys.-Usp. {\bf 39} 1193
  {(Translated from the Russian:} 
1996 {Usp. Fiz. Nauk} {\bf 166} 1257)

\bibitem{Vitalik} Antonov N V, Gulitskiy N M, Kakin P I and Serov V D 2021 {\it Effects of turbulent environment and random noise on self-organized critical behavior: Universality versus nonuniversality} Phys. Rev. E {\bf 103} 042106

\bibitem{AKa}
Antonov N V and Kapustin A S 2010 {\it Effects of turbulent mixing on critical behaviour in the presence of compressibility: Renormalization group analysis of two models} J. Phys. A: Math. Theor. {\bf 43} 405001

\bibitem{AntK}
Antonov N V and Kapustin A S 2012 {\it Critical behaviour of the randomly stirred dynamical Potts model: Novel universality class and effects of compressibility} J. Phys. A: Math. Theor. {\bf 45} 505001

\bibitem{Basu} Haldar A and Basu A 2022 {\it Disorders can induce continuously varying universal scaling in driven systems} Phys. Rev. E {\bf 105} 034104

\bibitem{Kohmoto} Kohmoto M, den Nijs M and Kadanoff L P 1981 {\it Hamiltonian studies of the $d=2$ Ashkin-Teller model} Phys. Rev. B {\bf 24} 5229

\bibitem{AV}
Antonov N V and Vasil'ev A N 1995 {\it The quantum-field renormalization group in the problem of a growing phase boundary} JETP {\bf 81} 485
(Translated from the Russian: ZhETF {\bf 108} 885)

\bibitem{JETP1}
Antonov N V 1997 {\it The renormalization group in the problem of turbulent convection of a passive scalar impurity with nonlinear diffusion} JETP {\bf 85} 898–906
(Translated from the Russian: 1997 ZhETF {\bf 112} 1649)

\bibitem{AK3} Antonov N V and Kakin P I 2017 {\it Scaling in erosion of landscapes: Renormalization group analysis of a model with infinitely many couplings} Theor. Math. Phys. {\bf 190}(2) 193-203

\bibitem{AK2} Antonov N V and Kakin P I 2017
{\it Scaling in erosion of landscapes: Renormalization group analysis of a model with turbulent mixing} J. Phys. A: Math. Theor. {\bf 50} 085002

\bibitem{ABK}
Antonov N V, Babakin A A and Kakin P I 2022 {\it Strongly Nonlinear Diffusion in Turbulent Environment: A Problem with Infinitely Many Couplings} Universe {\bf 8} 121
 
\end{thebibliography}
\end{document}